\documentstyle[epsf,epsfig,wrapfig,e-e-ijmpa]{article}

\begin{document}

\def\pole{{\cal P}_e}
\def\polp{{\cal P}_p}
\def\pollin{{\cal P}_e^{lin}}
\def\pollum{{\cal P}_e^{lum}}
\def\dpBMT{\Delta P^{BMT}}
\def\dplum{\Delta P_{IP}^{lum-wt}}
\def\dplumBMT{\Delta P_{IP}^{lum-wt,BMT}}
\def\dplumST{\Delta P_{IP}^{lum-wt,ST}}
\def\dpIPBMT{\Delta P_{IP}^{BMT}}
\def\dpIPST{\Delta P_{IP}^{ST}}
\def\dpIP{\Delta P_{IP}}
\def\dpCIP{\Delta P_{CIP}}
\def\dpCIPmeas{\Delta P_{CIP}^{meas}}
\def\dpCIPBMT{\Delta P_{CIP}^{BMT}}
\def\polg{{\cal P}_\gamma}
\def\rar{\rightarrow}
\def\lar{\leftarrow}
\def\epem{$e^+e^- \,$}
\def\emem{$e^-e^- \,$}
\def\alr{$A_{LR} \,$}
\def\alrm{A_{LR}}
\def\z0{$Z^0 \,$}
\def\sintw{\sin^2(\theta_W^{eff}) \,}
\def\pe{$P_e \,$}
\def\pelum{$P_e^{lum} \,$}
\def\ecm{E_{CM}}
\def\ecmlum{\langle E_{CM}^{lum-wt} \rangle}
\def\ecmbias{E_{CM}^{bias}}
\def\eav{\langle E \rangle }
\def\elumi{\langle E^{lum-wt} \rangle }
\def\einia{\langle E_1^{ini} \rangle }
\def\einib{\langle E_2^{ini} \rangle }
\def\eluma{\langle E_1^{lum-wt} \rangle }
\def\elumb{\langle E_2^{lum-wt} \rangle }

\markboth{M. Woods {\it et al.}}
{($L,E,P$) Studies}

\begin{flushright}
{\small
SLAC--PUB--10353\\
IPBI TN--2004--1 \\
February, 2004\\}
\end{flushright}
\vspace{.8cm}

\title{Luminosity, Energy and Polarization Studies for the Linear Collider: \\ Comparing \epem and \emem for NLC and TESLA
}

\author{\footnotesize M. Woods, K.C. Moffeit, T.O. Raubenheimer, A. Seryi, C. Sramek\footnote{
Supported by the Department of Energy, 
Contract DE-AC03-76SF00515.}}

\address{Stanford Linear Accelerator Center \\
Stanford University, Stanford, CA 94309
}

\author{A. Florimonte}

\address{Santa Cruz Institute for Particle Physics \\
University of California, Santa Cruz, CA 95064 
}

\maketitle


\begin{abstract}

We present results from luminosity, energy and polarization studies at a future Linear Collider.  We compare \epem and \emem modes of operation and consider both NLC and TESLA beam parameter specifications at a center-of-mass energy of 500 GeV.  Realistic colliding beam distributions are used, which include dynamic effects of the beam transport from the Damping Rings to the Interaction Point.  Beam-beam deflections scans and their impact for beam-based feedbacks are considered.  A transverse kink instability is studied, including its impact on determining the luminosity-weighted center-of-mass energy.  Polarimetry in the extraction line from the IP is presented, including results on beam distributions at the Compton IP and at the Compton detector.

\keywords{linear collider; luminosity spectrum; polarimetry.}
\end{abstract}

\vfill

\begin{center}
Presented at \\
\vspace{3mm}
{\it 
5th International Workshop on Electron-Electron Interactions at TeV Energies \\
December 12-14, 2003 \\
UC Santa Cruz, Santa Cruz, CA, USA}  \\
\end{center}

\pagebreak

\section{Introduction}	
We perform simulations of collisions at the Linear Collider (LC) Interaction Point (IP), and also of transport of the disrupted beams after collision from the IP to the beam dumps.  We compare \epem and \emem modes of operation and consider both NLC and TESLA beam parameter specifications at a center-of-mass energy of 500 GeV.  We use input files of the colliding beam distributions that were generated for the recent TRC study,\cite{trc} which compares the NLC and TESLA technical designs for a future Linear Collider.  The TRC beam parameter files were generated using a full beam transport simulation from the Damping Rings to the IP to achieve realistic colliding beam distributions.  These TRC files were used to create {\it electron.ini} and {\it positron.ini} files, which we use as input for a GUINEA-PIG\cite{schulte} simulation of beam-beam collisions.  The GUINEA-PIG (G-P) simulation is used to generate files of the outgoing beam distributions ({\it beam1.dat} and {\it beam2.dat}) to the extraction line.  It is also used to generate files ({\it lumi.dat}) of the luminosity-weighted beam energy distributions.  We use the same incoming beam parameters (the TRC distributions) for \epem and \emem collider modes.  These parameters are summarized in Table~\ref{tab:beamparams}.

The NLC extraction line design\cite{nosochkov1,nosochkov2} is used for beam transport from the IP to the beam dumps.  The extraction line performs two main functions, which are to provide beam diagnostics and to cleanly transport the disrupted beams (and secondary particles) to the beam dumps.  The diagnostics include a Compton polarimeter with the Compton Interaction Point (CIP) located in the middle of a chicane, approximately 60 meters downstream of the Linear Collider IP.\cite{woods1,rowson}  A GEANT-3 simulation of the extraction line\cite{maruyama} is used; its results have been  checked against a DIMAD simulation\cite{nosochkov2} and found to agree.

Many studies on the physics impact of beam-beam collisions exist for both the NLC and TESLA beam parameters at a 500-GeV Linear Collider.  But for the large majority of these, only one machine design is studied; and the studies do not include realistic colliding beam distributions.  Typically, gaussian beam distributions are assumed with no correlations between the beam parameters.  Our study uses more realistic colliding beam distributions, including effects from Linac wakefields and beam component misalignments.  (The TRC study, that generated our input colliding beam distributions, applied a set of misalignment errors to the Linac and then used simple Linac steering and a model of IP feedback to optimize the beam transport, center the colliding beams and optimize luminosity.)  Our studies consider several different {\it machines} (with different sets of initial misalignments) by using several sets of TRC beam files.  We directly compare the baseline NLC and TESLA beam parameters, for both \epem and \emem collider modes.   

\begin{table} [tbp]
\caption{NLC and TESLA beam parameters for a 500 GeV Linear Collider.}
\vspace{2mm}
{\begin{tabular}{@{}|c|c|c|@{}} 
\hline
{\bf Beam Parameter}		& {\bf NLC-500} 		& {\bf TESLA-500}	\\
\hline
Beam Energy 		& 250 GeV			& 250 GeV			\\
Repetition Rate		& 120 Hz			& 5 Hz 			\\
Bunch Charge		& $0.75 \cdot 10^{10}$	&  $2.0 \cdot 10^{10}$	\\
Bunches per rf pulse	&  192			&  2820			\\
Bunch spacing		& 1.4 ns			&  337 ns			\\

$\gamma \epsilon_x, \gamma \epsilon_y$ & $(360, 4.0) \cdot 10^{-8}$ m-rad		&  $(1000, 3.0) \cdot 10^{-8}$ m-rad						\\
$\beta_x, \beta_y$	& (8,0.11) mm		&  (15, 0.40) mm		\\
$\sigma_x, \sigma_y$	& (243,3.0) nm		&  (554,5.0) nm		\\
$\sigma_z$			& 110 $\mu$m		&  300 $\mu$m		\\
$\frac{\sigma_E}{E}$ electrons & 0.30$\%$		&  0.14$\%$ 		\\
$\frac{\sigma_E}{E}$ positrons & 0.30$\%$		&  0.07$\%$ 		\\
Geometric Luminosity	&  $1.4 \cdot 10^{34} cm^{-2}s^{-1}$ & 
$1.6 \cdot 10^{34} cm^{-2}s^{-1}$ \\
\hline
\end{tabular}}
\label{tab:beamparams}
\end{table}

\section{Luminosity Studies}
The geometric luminosity, $L_0$, at the Linear Collider is given by
\begin{equation}
L_0 = \frac{f_{rep}N^2}{4\pi\sigma_x\sigma_y},
\label{eq:lumi}
\end{equation}
where $f_{rep}$ is the number of colliding bunches per second; $N$ is the bunch charge; and $\sigma_x,\sigma_y$ are the horizontal and vertical beam sizes.  Beam-beam focusing effects can enhance $L_0$ for \epem collisions.  For \emem collisions, however, there is an anti-pinch effect and the luminosity is reduced.   This pinch (or anti-pinch) effect can be expressed as
\begin{equation}
L=L_0 \cdot H_D,
\label{eq:lumipinch}
\end{equation}
where $H_D$ is the pinch enhancement factor.  For the NLC-500 and TESLA-500 beam parameters the horizontal pinch is negligible, while the vertical pinch is significant.  The pinch effect also results in decreased (increased) sensitivity of the luminosity, $L$, to vertical offsets of the \epem (\emem) colliding beams.

\begin{figure}
\begin{center}
\epsfig{file=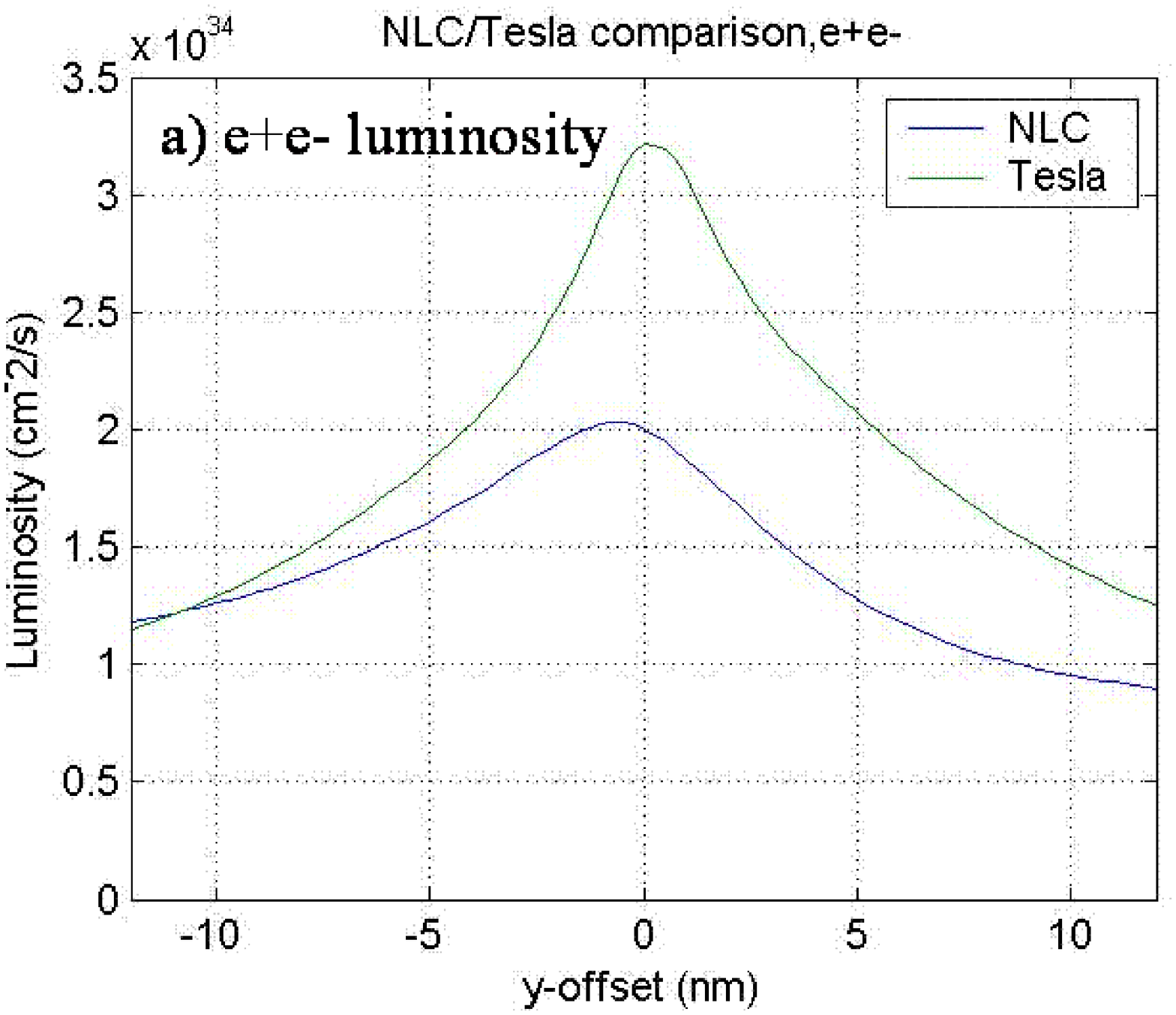,width=6.2cm}
\epsfig{file=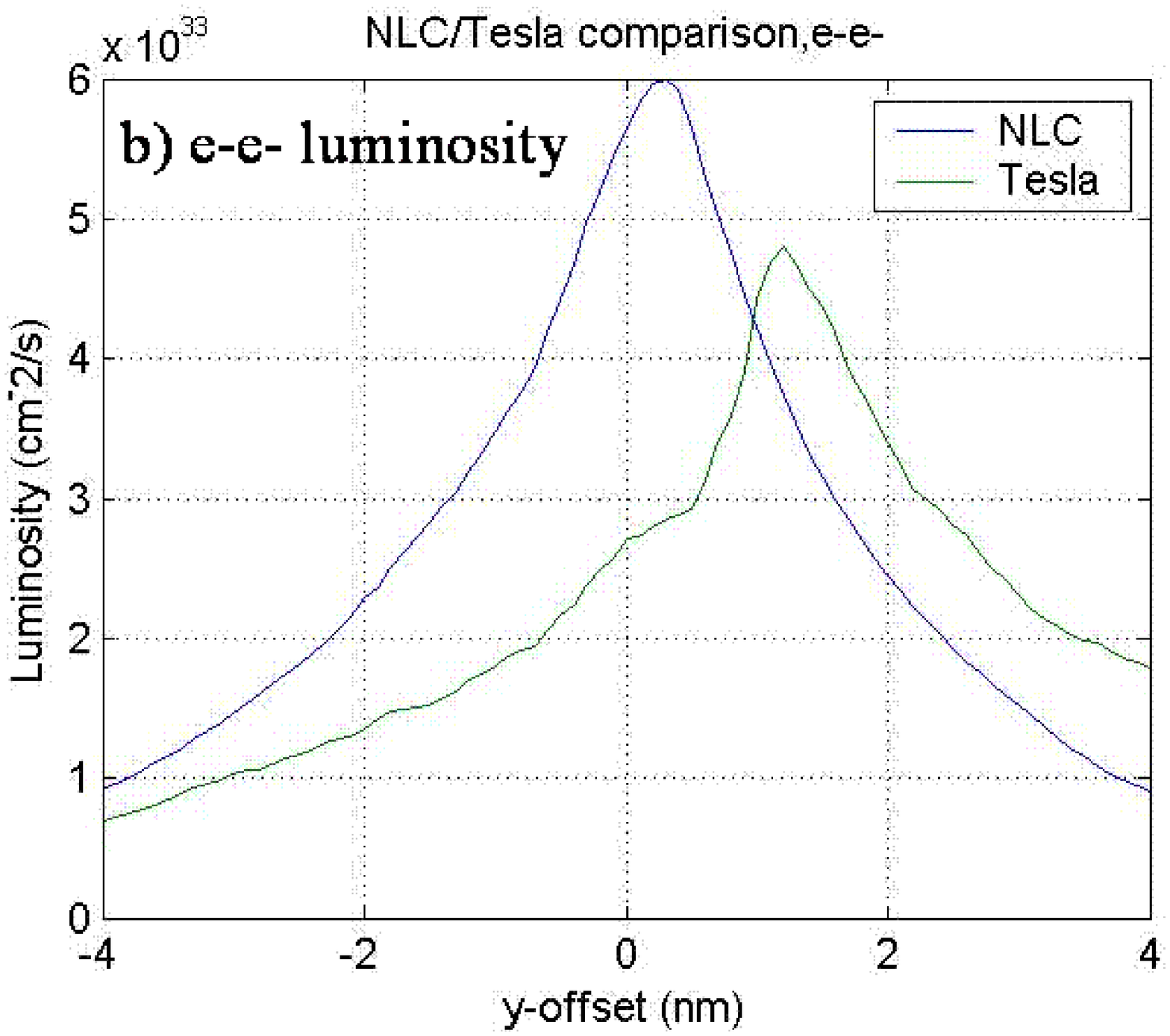,width=6.2cm}
\end{center}
\caption{The luminosity versus vertical offset is plotted for a) \epem collisions and b) \emem collisions, for one set of TRC files.  The TESLA beam parameters result in a higher luminosity for \epem, but lower luminosity for \emem.}
\label{fig:lumi}
\end{figure}

The luminosity as a function of the vertical offset between the colliding beams at a 500 GeV Linear Collider is plotted in Figure~\ref{fig:lumi}, for one set of TRC files.  (The detailed results vary with the files used, but the plots shown in Figure~\ref{fig:lumi} give a good representation of the general features observed.)  We observe higher luminosity with TESLA parameters for \epem collisions, while NLC parameters give higher luminosity for \emem collisions.  The geometric luminosity, $L_0$, is roughly equal in the two designs.  But the TESLA beam parameters give $H_D=2.1$ for \epem collisions and $H_D=0.3$ for \emem collisions, while NLC beam parameters give $H_D=1.5$ for \epem and $H_D=0.4$ for \emem.  

\begin{figure}
\begin{center}
\epsfig{file=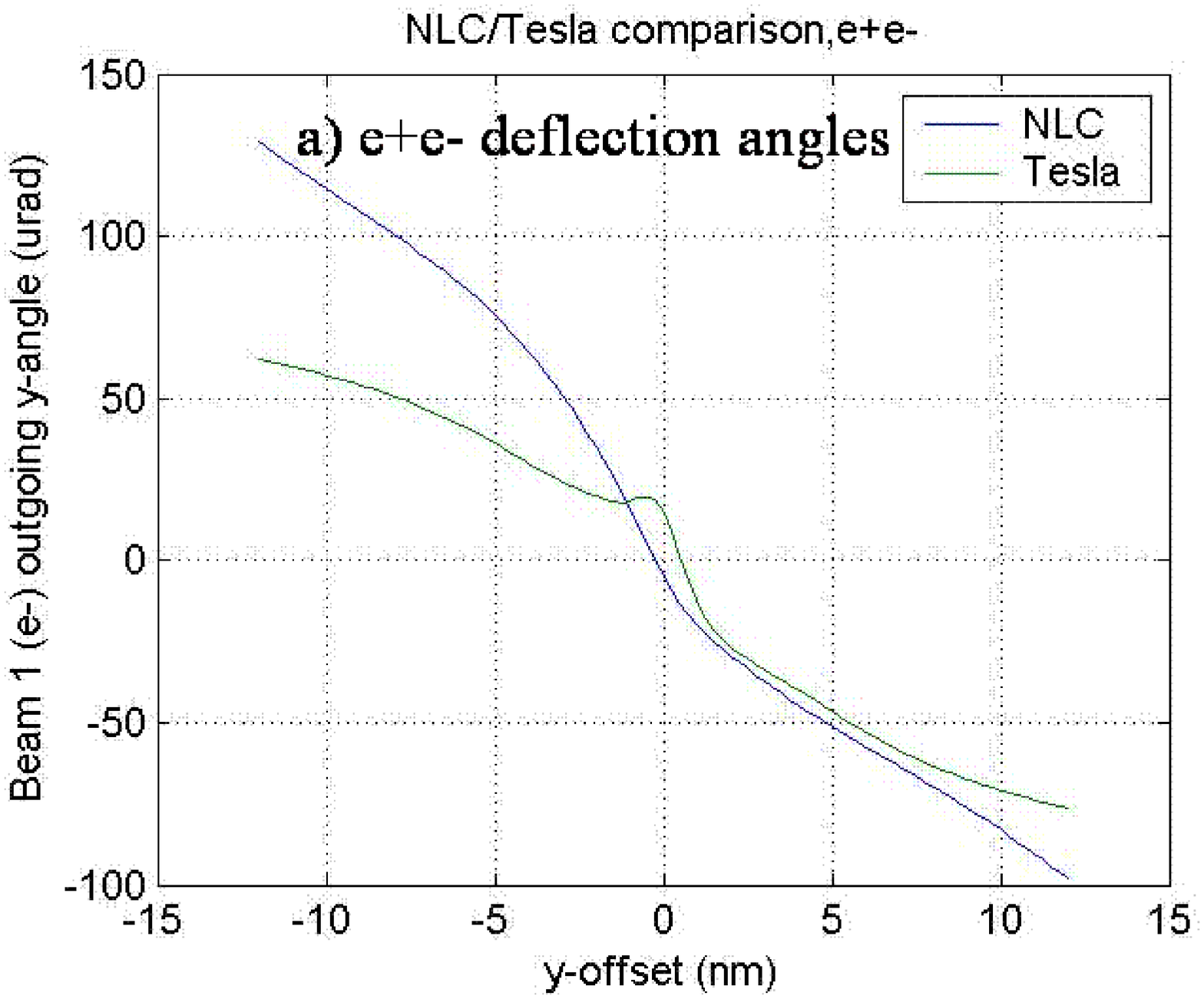,width=6.2cm}
\epsfig{file=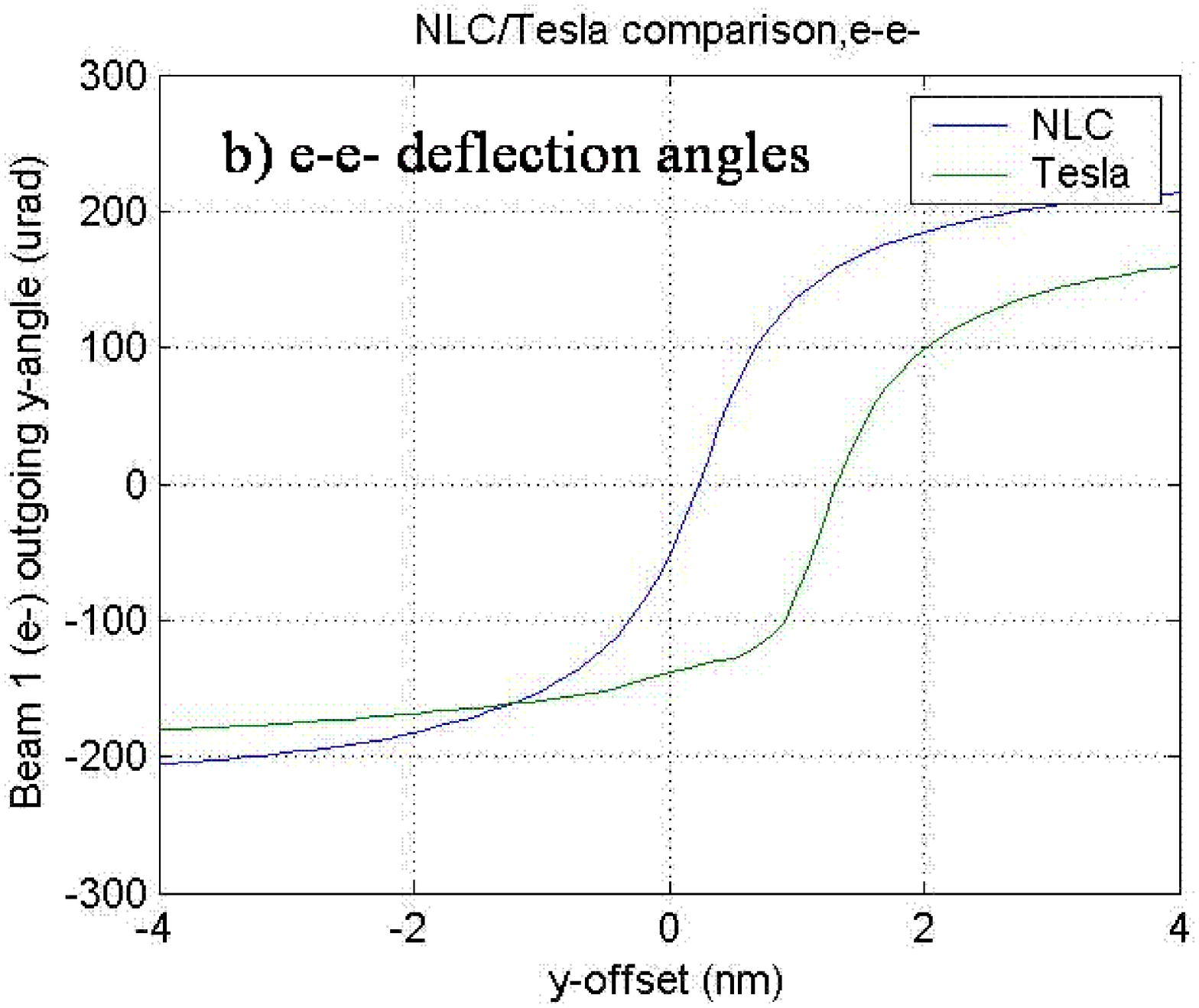,width=6.2cm}
\end{center}
\caption{The vertical deflection angles are plotted versus vertical offset for a)\epem and b) \emem collisions, for one set of TRC files.  The NLC beam parameters result in larger vertical deflection angles.}
\label{fig:ydefl}
\end{figure}

One feature we observe in Figure~\ref{fig:lumi} is that the maximum luminosity may not occur at zero offset.  This effect is more enhanced for \emem collisions.  To examine this further, we plot in Figure~\ref{fig:ydefl} the average beam-beam deflection angle as a function of vertical offset.  (The same TRC files are used here as for the results shown in Figure~\ref{fig:lumi}.  Again, the detailed results vary with the input TRC files used, but the plots shown give a good representation of the general features observed.)  We find that the deflection angles can also be non-zero at zero offset, and again the effect is enhanced for \emem collisions.  We observe, however, that maximum luminosity is achieved when the outgoing deflection angles are approximately zero, rather than when the colliding beams have zero offset.  Thus beam-beam deflection feedbacks, which are planned to optimize and stabilize luminosity at both NLC and TESLA, should still work.

The source of reduced luminosity and significant deflection angles when the beams are centered can be traced to a vertical {\it kink instability}\cite{kink1,kink2} and the use of realistic colliding beam distributions at the IP.\cite{brinkman}  When we simulate collisions of beams with gaussian beam parameter distributions and no correlations among the beam parameters, we find that maximum luminosity and zero deflection angles both occur when the beam offsets are zero.  The {\it kink instability} is discussed in more detail in the following section, where we consider its impact on determining the luminosity-weighted center-of-mass energy.  

We also observe that the luminosity and deflection angle curves in Figures~\ref{fig:lumi} and~\ref{fig:ydefl} are significantly narrower for \emem collisions than for \epem collisions.  This presents a significant difficulty for beam-based feedbacks to stabilize collisions and optimize luminosity.  One study for the TESLA-500 machine design has considered the narrow \emem deflection curve and how well a fast intra-train feedback can be expected to work to keep the beams centered.\cite{schreiber}   This feedback takes advantage of the long (0.95 ms) bunch train and uses the observed deflection angles at the head of the train to make corrections to bring the rest of the train into collision.  If the bunch-to-bunch jitter within a train is sufficiently small, the study showed that this feedback can work with only a small loss in luminosity.  For the NLC machine design, however, the bunch train is short (270 ns) and the expected latency of the feedback (less than 60 ns)\cite{font} would constitute a non-negligible fraction of the train.  An intra-train deflection feedback for \epem collisions with the NLC-500 parameters can still be expected to work reasonably well if needed (with residual train-to-train jitter offsets of 5-10 nm or less).  A slower inter-train feedback is significantly easier and will work well for NLC-500 \epem collisions.  It is easier than the TESLA intra-train feedback because of the larger deflection angles, the more well-behaved deflection curve and the larger 8-ms sampling interval to calculate and apply correction signals.  The inter-train sampling rate of 120 Hz for NLC is much better than the 5 Hz inter-train sampling rate for TESLA; it is lower, though, than TESLA$'$s average 14 kHz sampling rate (3 MHz instantaneous sampling rate) that is possible if TESLA intra-train sampling works well.  

For \emem collisions, the NLC deflection curve shown in Figure~\ref{fig:ydefl}b may be too narrow to expect either intra-train or inter-train beam-beam deflection feedbacks to perform well.  One recent study\cite{sramek} considered this and suggested increasing the vertical spotsize (by increasing the vertical beta function) for \emem collisions by a factor 7 to achieve a deflection curve similar to that for \epem collisions.  The resulting luminosity (for NLC) was only about $10\%$ of the \epem luminosity.  As experience is gained in machine operation, however, narrower deflection curves than the one plotted in Figure~\ref{fig:ydefl}a for NLC-500 may be tolerated.  In the case of the TESLA beam parameters, the \emem luminosity is observed in our study to be $\approx 15\%$ of the \epem luminosity before considering the difficulties the narrow deflection scans present to the beam-based feedbacks.  

Realistic colliding beam parameters and beam-based feedbacks are likely to yield \emem luminosities of order $10-20\%$ of \epem luminosities.  This range of results should be similar for the NLC and TESLA designs.  This reduced \emem luminosity is significantly lower than the canonical factor of 1/3 that typically appears in the literature.

\section{Energy Studies}

Energy and luminosity are the most important parameters characterizing a future Linear Collider and determining its reach for new physics.  In addition, the LC has the capability for many precision measurements, which further extend this reach.  Electron(-positron) colliders can have a distinct advantage over proton colliders because of the well defined initial state.  However, beam energy spread, beamsstrahlung and beam disruption angles make the colliding beam parameters less precise.  This can reduce the achievable precision for some measurements.  In this paper we consider the effect of the beam energy spread on the precision with which one can determine the average luminosity-weighted center-of-mass energy, $\ecmlum$.

The LC will precisely measure the top quark mass and the Higgs boson mass (if the Higgs boson exists).  These measurements motivate determining $\ecmlum$ to 200 parts per million (ppm) or better.\cite{ipbiwhite}  Improving the current W mass measurement or performing a very precise \alr measurement in a "Giga-Z" program motivate determining $\ecmlum$ to 50 ppm or better.\cite{ipbiwhite}

The beam energy spectrometers measure the average beam energy, $\eav$, which can differ from $\elumi$ due to effects from beam energy spread and beamsstrahlung.  The 50-200 ppm desired precision for $\ecmlum$ is well below the 3000 (1000) ppm rms energy spread for NLC-500 (TESLA-500).  Beam parameter correlations and aberrations can cause the luminosity to vary over the phase space of the incoming beam parameters, and can lead to
\begin{equation}
\eav \neq \elumi.
\end{equation}

In this section we consider one effect that causes a bias to the $\ecmlum$ determination, which arises from three ingredients:  beam energy spread, energy-z correlation within a bunch and a y-z kink instability.  The bias assumes a certain analysis technique, which is currently favored for determining $\ecmlum$.  This analysis uses beam energy spectrometer measurements to measure the average incident beam energies per bunch and it uses the acollinearity in Bhabha events to infer the effects from beam energy spread and beamsstrahlung.\cite{miller}  We do not consider here the potentially larger effect from beamsstrahlung.  (Beamsstrahlung induces an energy spread in $\ecmlum$  greater than the incoming beam energy spread does.  Our study effectively assumes that beamsstrahlung effects can be corrected for exactly using the Bhabha acollinearity analysis.)  The y-z {\it kink instability} has been considered in previous studies\cite{brinkman,schulte1,schulte2} for its effect on the luminosity at TESLA.  We present results here for its effect on the $\ecmlum$ determination.

\begin{figure}
\begin{center}
\epsfig{file=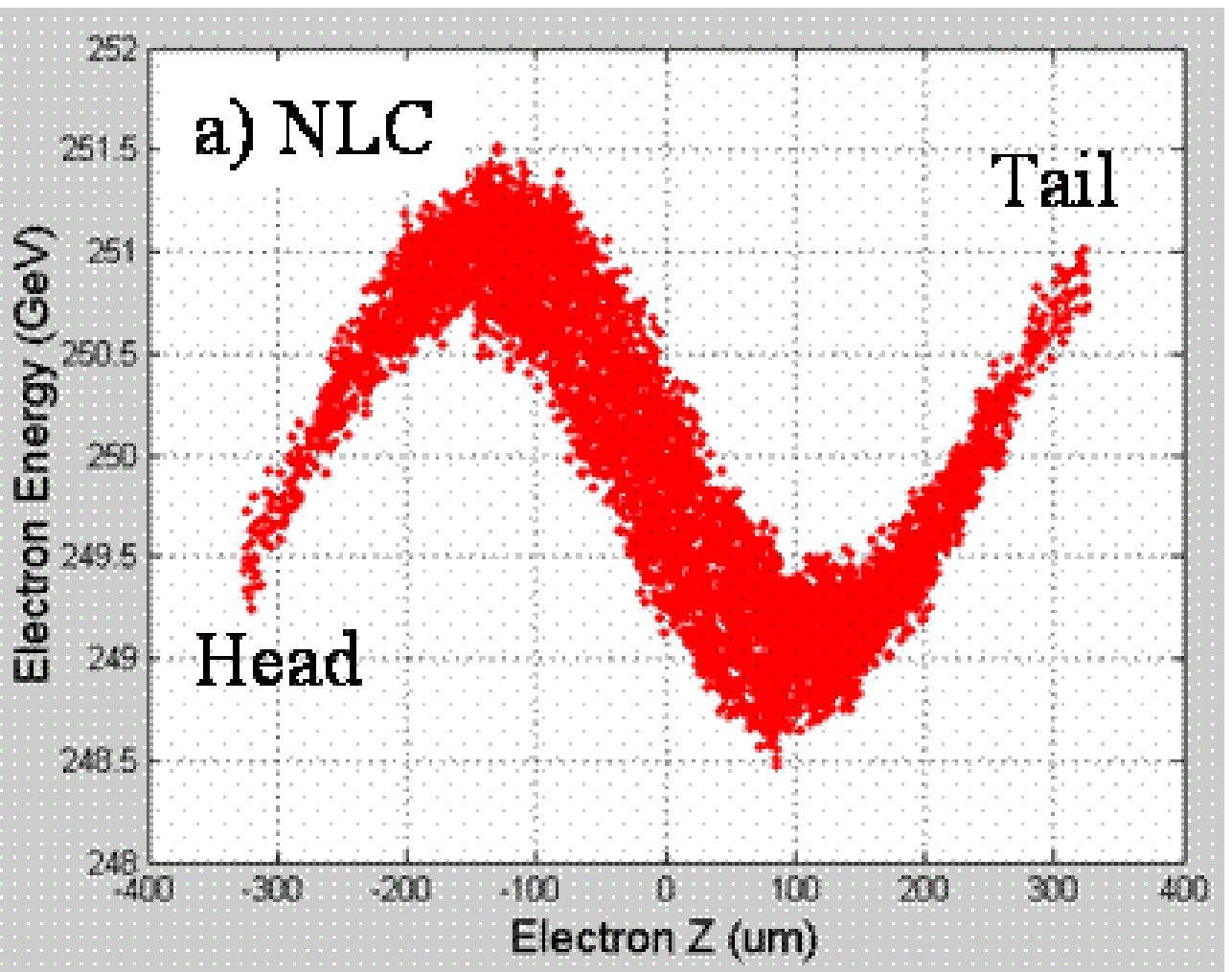,width=6.2cm}
\epsfig{file=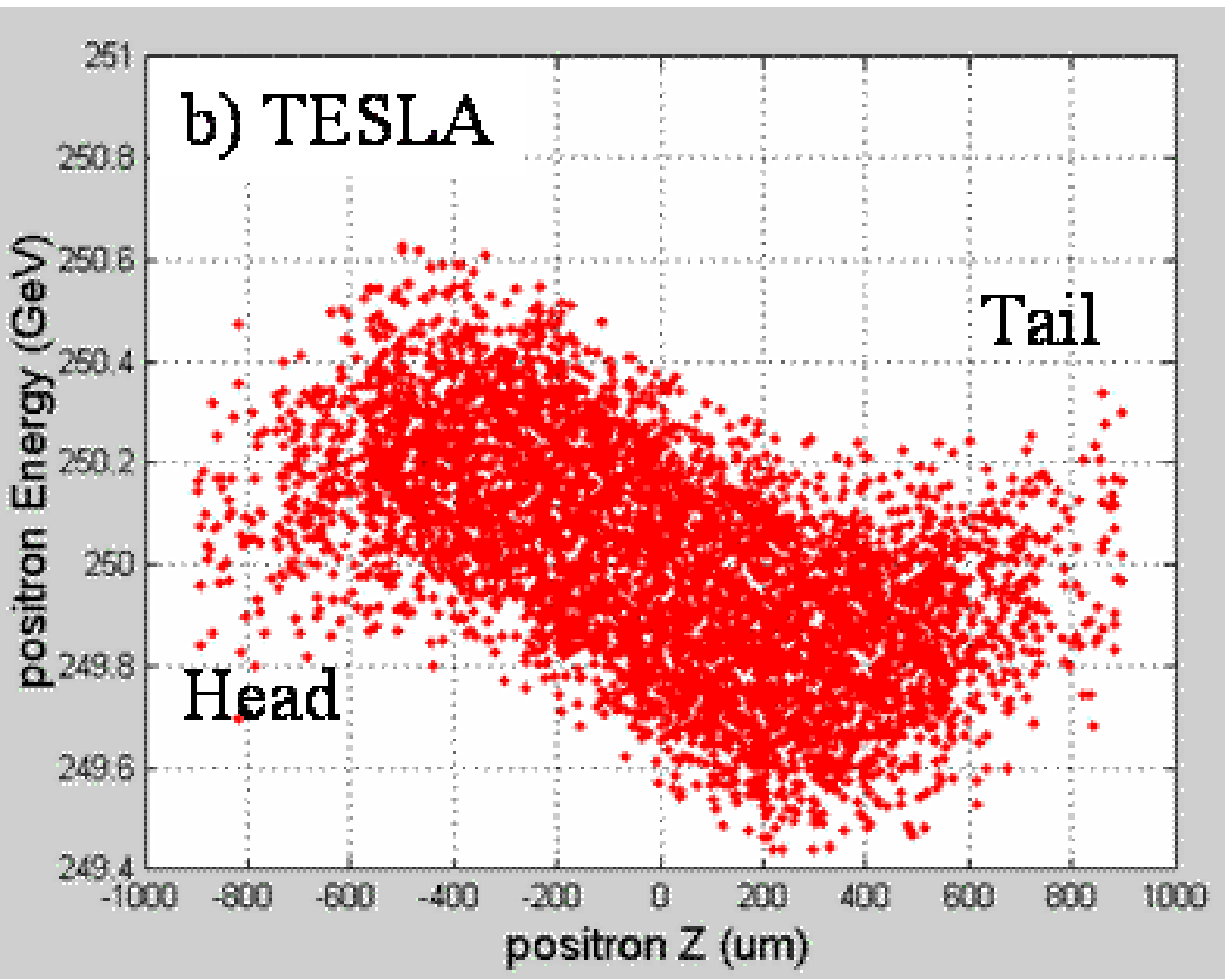,width=6.2cm}
\end{center}
\caption{The energy-z correlation incoming beams at the IP for a) electrons at NLC, and b) positrons at TESLA.} 
\label{fig:energy-z}
\end{figure}

We use the G-P simulation of beam-beam effects with the flags for beamsstrahlung and initial state radiation turned off.  We define $\ecmlum$ and the bias in determining it by, 
\begin{eqnarray}
\ecmlum  & = & \eluma + \elumb \\
\ecmbias & = & \ecmlum - \left( \einia + \einib \right),
\end{eqnarray}
where $\einia$ and $\einib$ are the average energies of the incoming beam distributions in the G-P files {\it electron.ini} and {\it positron.ini}; and $\eluma$ and $\elumb$ are the average energies of colliding particles that make luminosity, taken from the G-P file {\it lumi.dat}.  The energy spread and energy-z correlations of the incoming bunches to the IP are plotted in Figure~\ref{fig:energy-z}a) for NLC-500 electrons and in Figure~\ref{fig:energy-z}b) for TESLA-500 positrons.  NLC-500 has similar distributions for electrons and positrons, while the electron energy distribution for TESLA-500 has additional broadening (compared to TESLA-500 positrons) due to the undulator used for positron production.  The beam energy spread consists of two components: an uncorrelated contribution from the bunch compressors and the positron source undulator, and a correlated contribution from the residual of the BNS-damping energy spread.\cite{trc-espread}  The energy-z correlation arises from the need for BNS damping to prevent jitter amplification due to the transverse wakefields.  Because of the weaker wakefields in the superconducting TESLA design, the required BNS energy spread is about a factor of five smaller than in the normal conducting NLC design.

\begin{figure}
\begin{center}
\epsfig{file=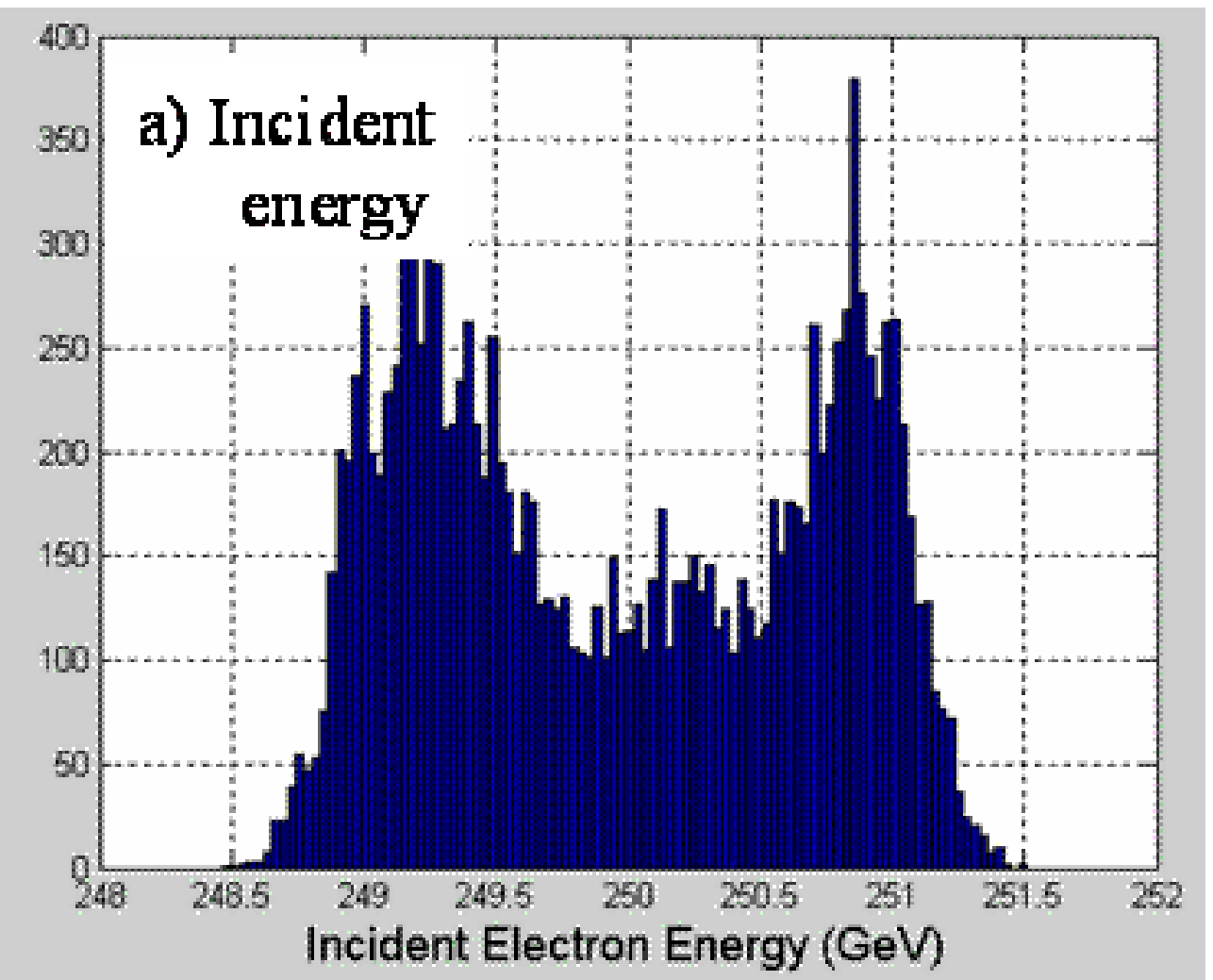,width=6.2cm}
\epsfig{file=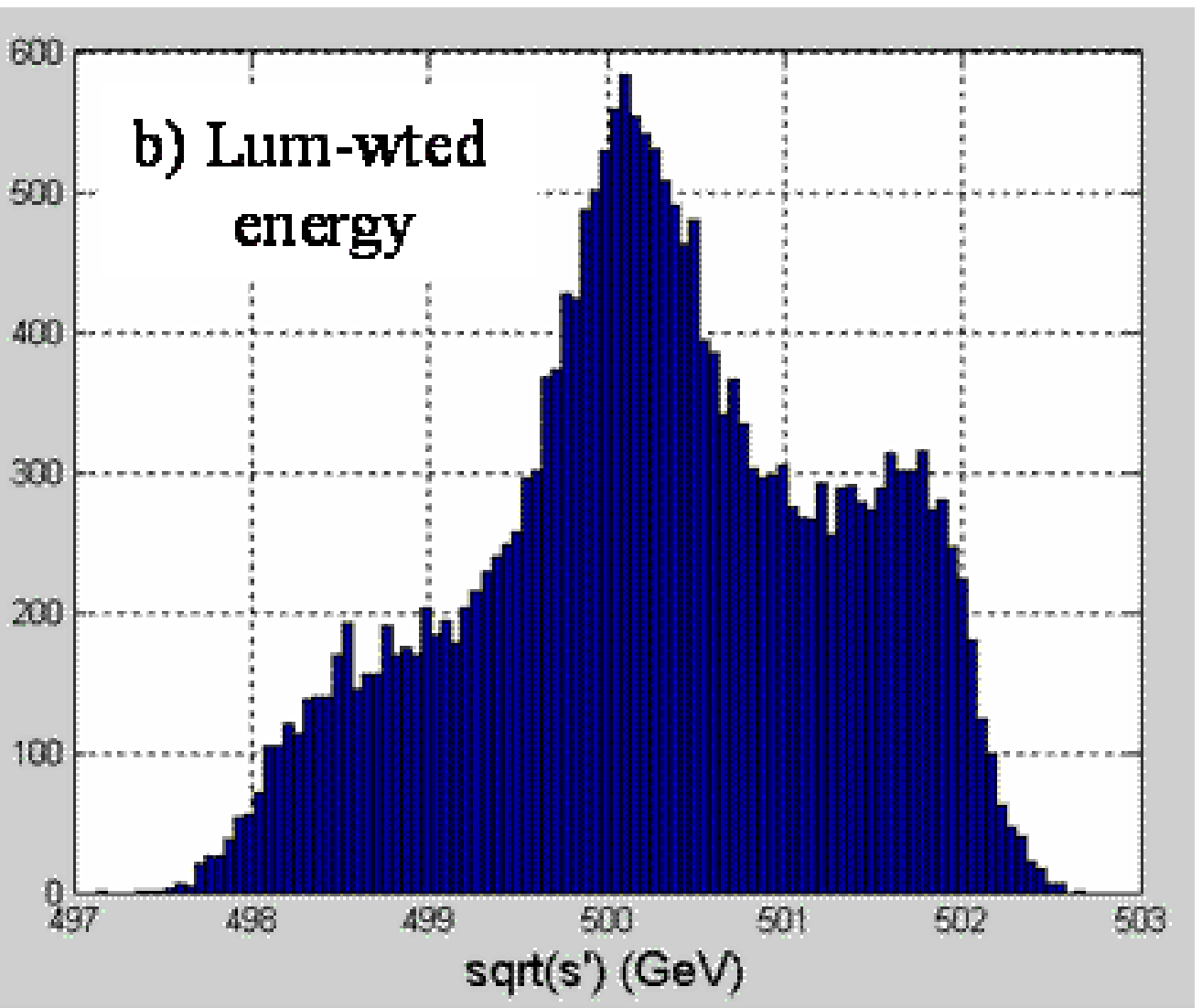,width=6.2cm}
\end{center}
\caption{The incident beam energy distribution (a) and the luminosity-weighted center-of-mass energy (b) for \epem collisions at NLC, for one set of TRC files.}
\label{fig:energyhistonlc}
\end{figure}

\begin{figure}
\begin{center}
\epsfig{file=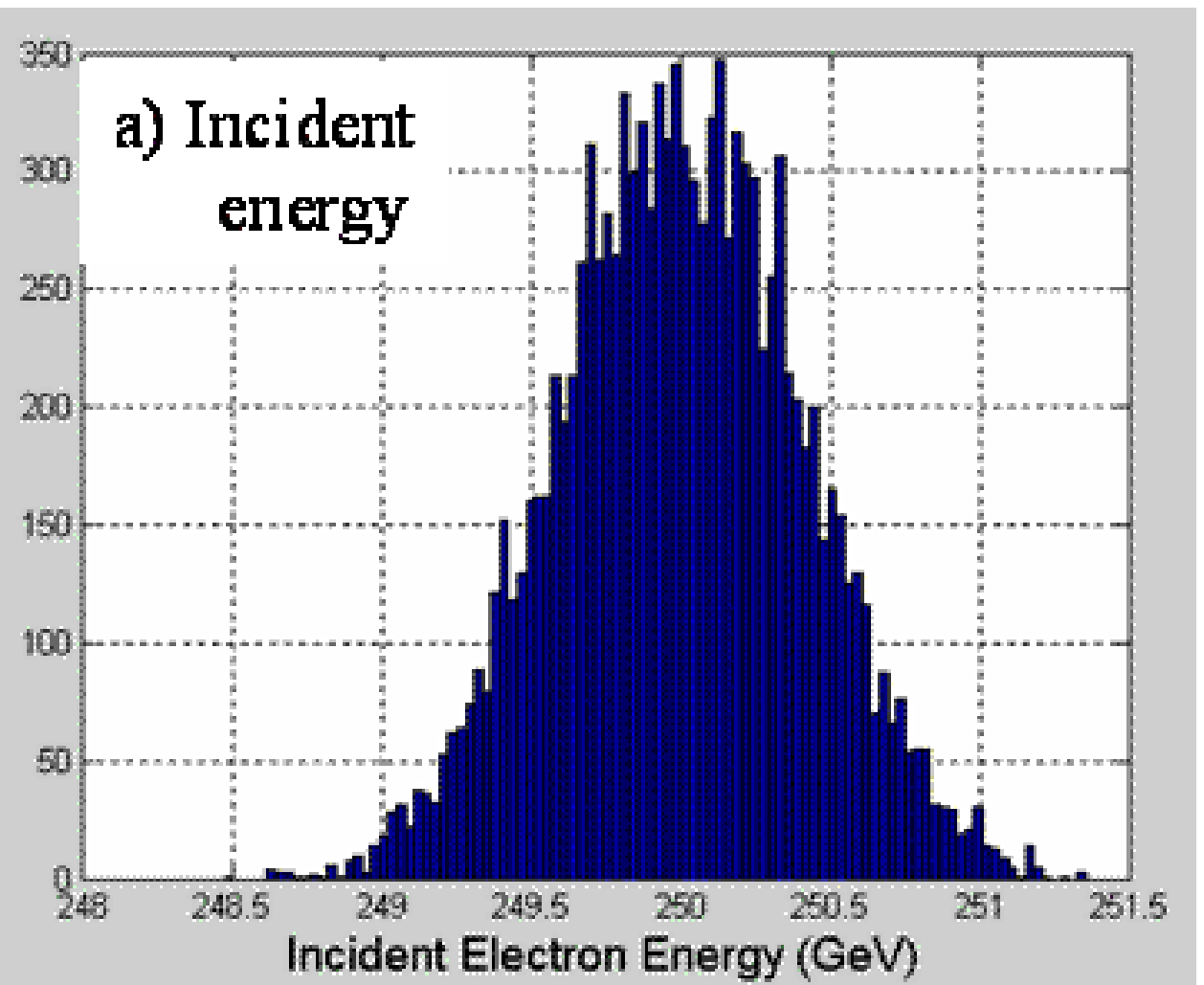,width=6.2cm}
\epsfig{file=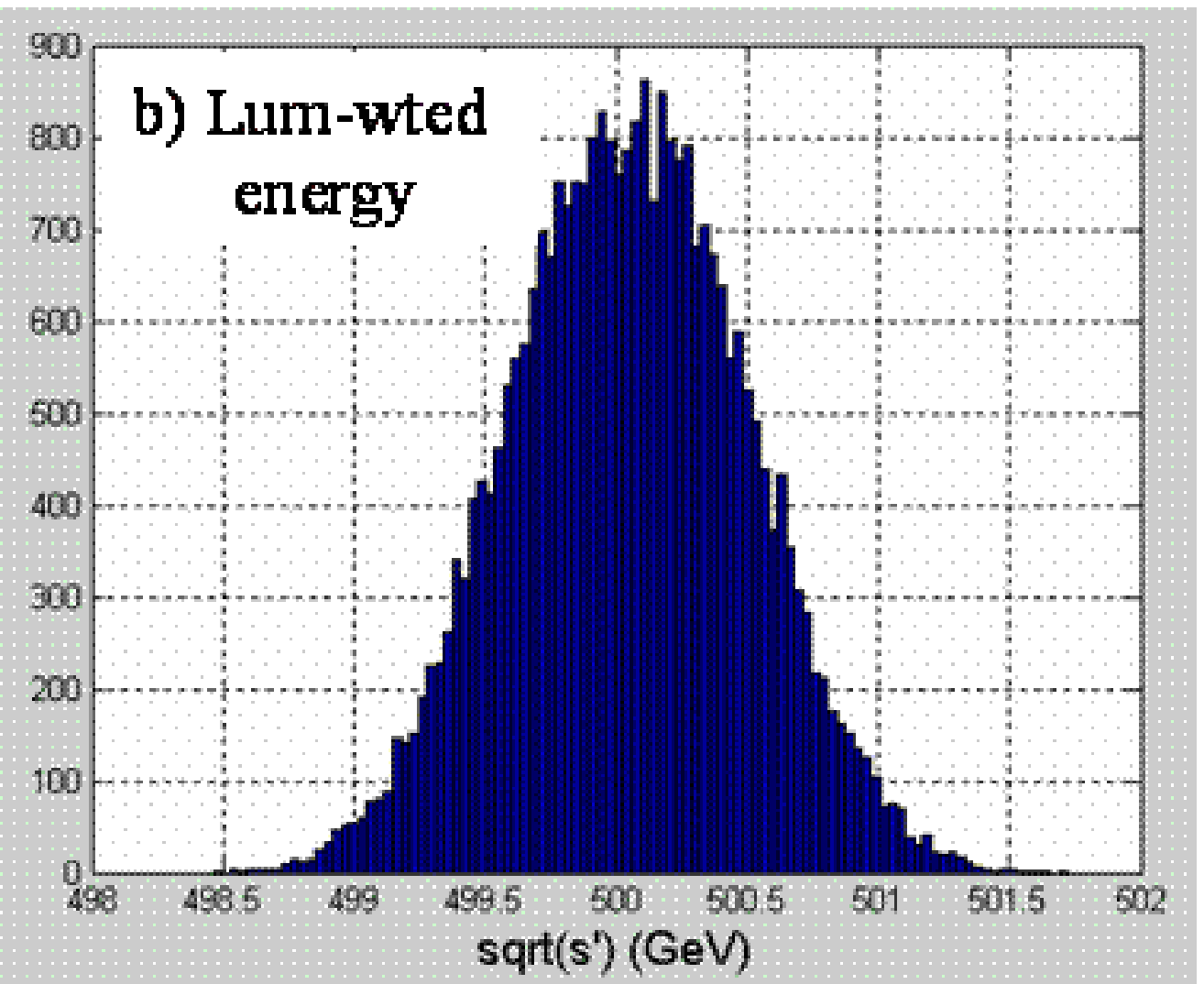,width=6.2cm}
\end{center}
\caption{The incident beam energy distribution (a) and the luminosity-weighted center-of-mass energy (b) for \epem collisions at TESLA, for one set of TRC files.}
\label{fig:energyhistotesla}
\end{figure}

For \epem collisions, histograms of the incident beam energy distributions and the G-P results for the $\ecmlum$ distribution are shown in Figure~\ref{fig:energyhistonlc} for NLC-500 and in Figure~\ref{fig:energyhistotesla} for TESLA-500.  The NLC-500 $\ecmlum$ distribution is clearly asymmetric.  This results from a kink instability that causes the heads of the bunches to have higher luminosity than the tails.  The colliding bunches can be viewed as long thin ribbons a few nanometers high, a few hundred nanometers wide and over 100 microns long.  Ideally the transverse distortions of these colliding ribbons is small along their longitudinal {\it z}-axis.  However, there can be small distortions in the incoming beams and additional distortions from beam-beam effects.  These are dynamic and evolve during the beam-beam collision.  When the vertical disruption gets large enough, a kink instability develops which can cause luminosity loss and, if there is an energy-z correlation, a significant ($>100$ ppm) $\ecmbias$ as well.

The $\ecmbias$ corresponding to the NLC-500 distributions shown in Figure~\ref{fig:energyhistonlc} is +550 ppm, and for the TESLA-500 distributions shown in Figure~\ref{fig:energyhistotesla} it is +95 ppm.  These results correspond to just one of the TRC files generated for each of NLC-500 and TESLA-500.  We have repeated this analysis for 6 different ({\it electron.ini} and {\it positron.ini}) TRC files for each of NLC-500 and TESLA-500.  These correspond to 6 sets of (random) misalignments, which, when combined with  simple beam tuning algorithms, achieve nominal luminosity with centered colliding beams.  (This mimics effects of larger errors that may be encountered, but also corrected with more sophisticated beam tuning algorithms.)  The variations in $\ecmbias$ observed for 6 sets of TRC input files are summarized in Table~\ref{tab:ecmbias}.

\begin{figure}
\begin{center}
\epsfig{file=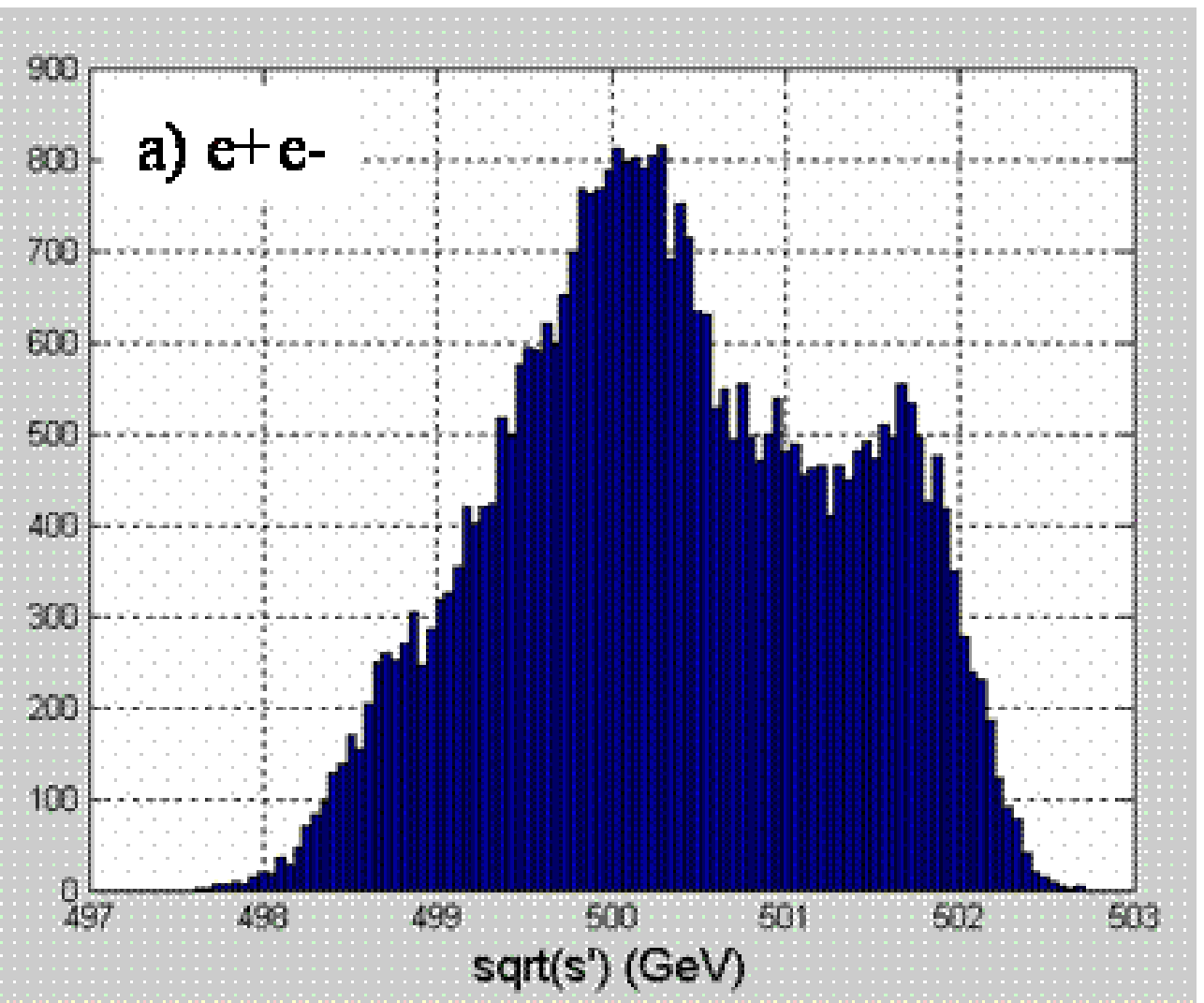,width=6.2cm}
\epsfig{file=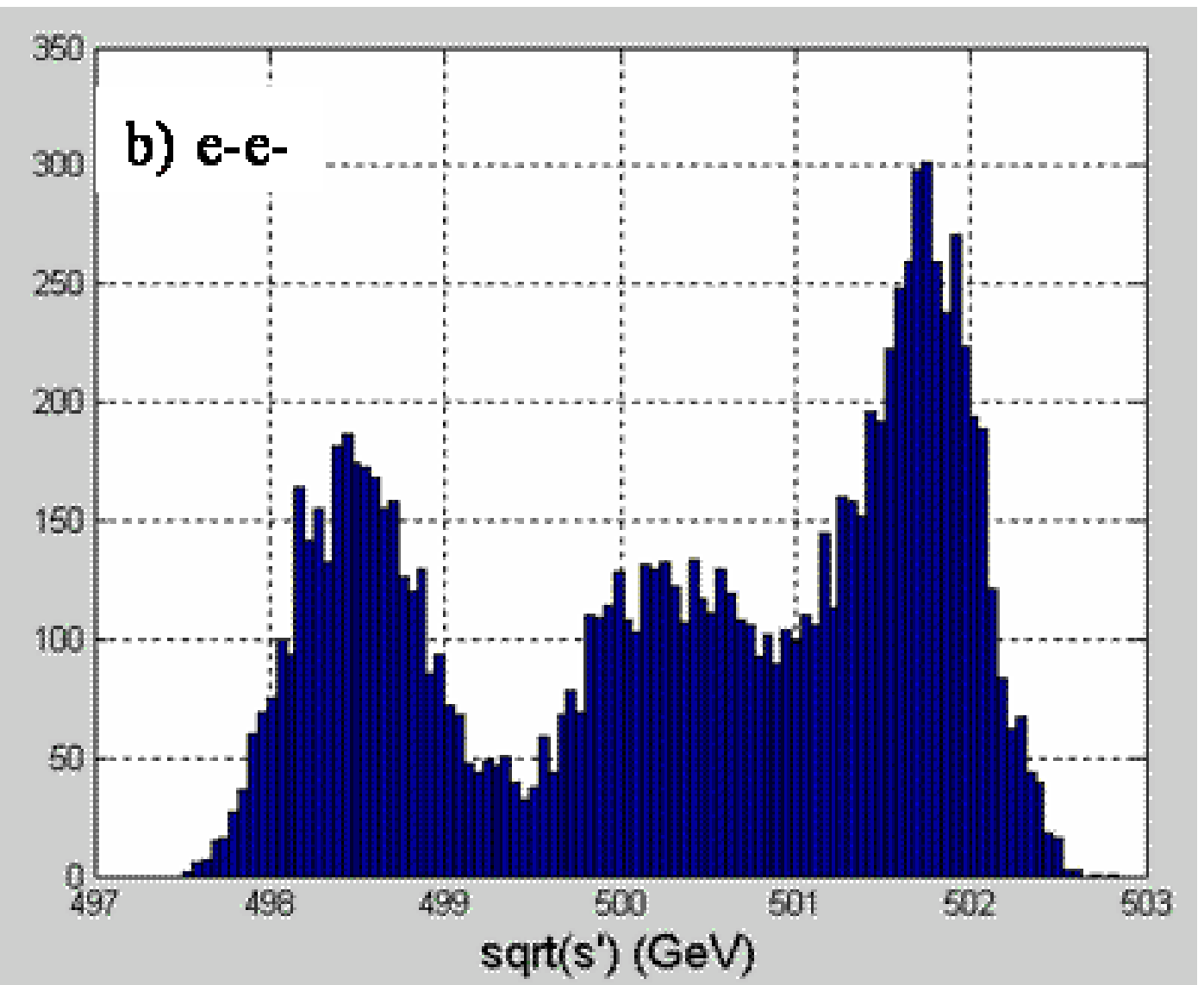,width=6.2cm}
\end{center}
\caption{The luminosity-weighted center-of-mass energy for one set of the NLC beam parameters, for both a) \epem and b) \emem collisions.}
\label{fig:lumwted_e_nlc}
\end{figure}

We have also examined the magnitude of $\ecmbias$ due to the kink instability for \emem collisions.  Figure~\ref{fig:lumwted_e_nlc} shows the luminosity-weighted $\ecm$ distribution for both \epem and \emem collisions, using the NLC-500 simulation file which exhibited the largest $\ecmbias$ for \epem collisions.  In this case, we find for \epem collisions $\ecmbias = +720$ ppm, and for \emem collisions $\ecmbias = + 690$ ppm.  The variations in $\ecmbias$ observed for \emem collisions are also summarized in Table~\ref{tab:ecmbias}.

\begin{figure}
\begin{center}
\epsfig{file=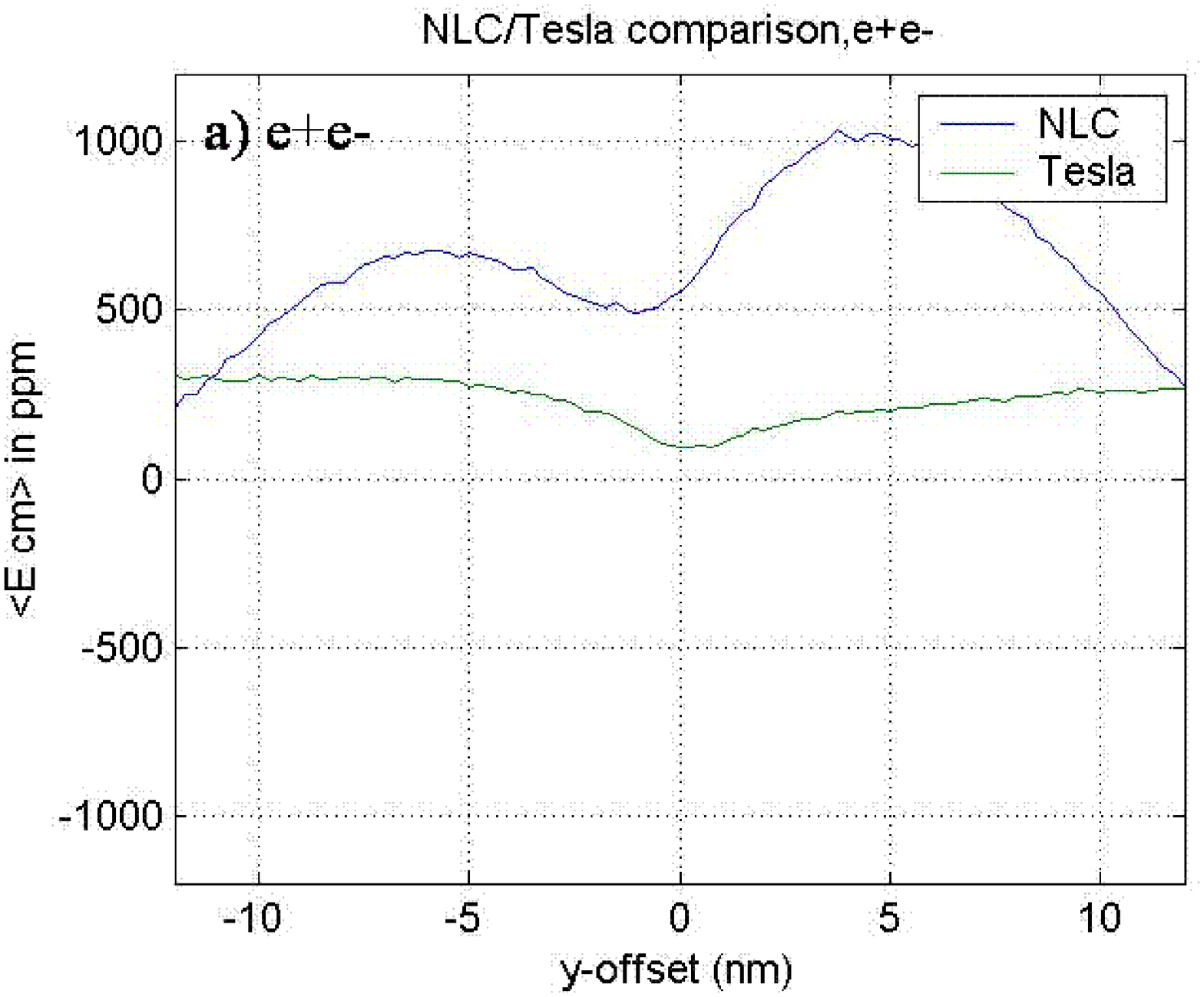,width=6.2cm}
\epsfig{file=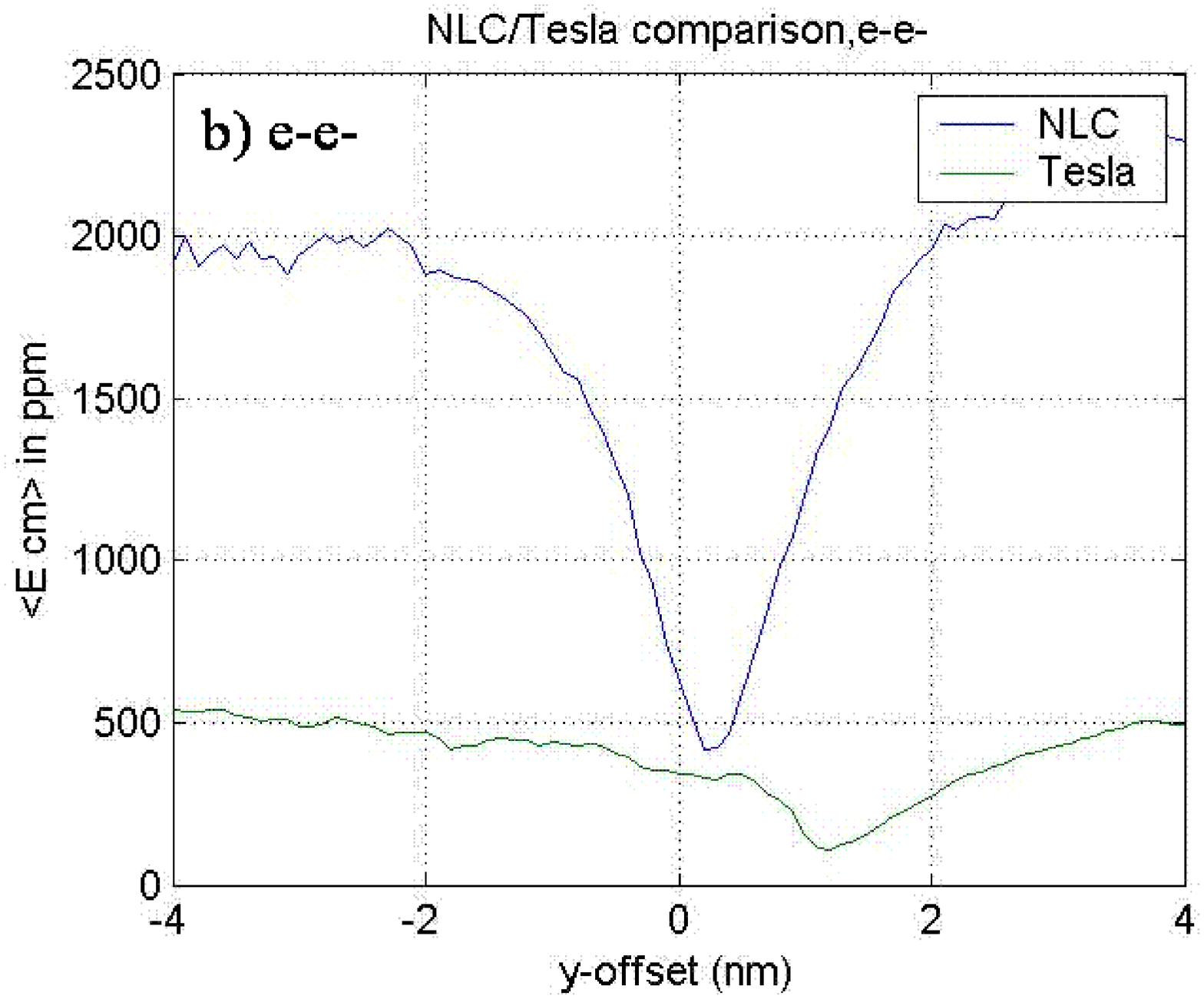,width=6.2cm}
\end{center}
\caption{The {\it bias} in the luminosity-weighted center-of-mass energy is plotted versus the vertical offset of the colliding beams for a) \epem collisions and b) \emem collisions.  The bias is larger for NLC than for TESLA.}
\label{fig:ecmbias_yoff}
\end{figure}

We have investigated how $\ecmbias$ may have additional dependence on aberrations at the IP due to residual beam position offsets, waist offsets and dispersion.  The results for vertical position offsets are shown in Figure~\ref{fig:ecmbias_yoff} for one of the NLC-500 and TESLA-500 simulations for both \epem and \emem collision modes.  (The details observed in these plots vary for different TRC files, but the results shown indicate the relevant features.)  The maximum size of effects we see for $\ecmbias$ from these studies are summarized in Table~\ref{tab:ecmbias}.  

We note that the $\ecmbias$ considered in this paper is due to one effect (energy spread and kink instability) and assuming one analysis technique (energy spectrometers and Bhabha acollinearity).  The study of other effects, such as beamstrahlung and disruption angles, is still in progress.  Incorporating realistic colliding beam distributions and investigating effects from asymmetric beamsstrahlung emission of the two colliding beams have not yet been done.  

We are pursuing other physics analyses\cite{ipbiwhite} (ex. $\gamma${\it Z}, {\it ZZ} and {\it WW} events), where we can utilize existing measurements of the {\it Z} and {\it W} masses.  We expect these analyses will resolve the $\ecmbias$ (apparent in the Bhabha acollinearity analysis method) to achieve $<200$ ppm precision in $\ecmlum$ for both NLC-500 and TESLA-500 machines.  However, the energy spread/kink instability effect we have discussed in this paper is a significant issue for achieving $<50$ ppm precision on $\ecmlum$ as desired for improved W-mass measurements, and also for achieving the best possible \alr measurement at Giga-Z (if polarized positrons or sub-0.1$\%$ polarimetery are available).  This is an important issue for both the NLC-500 and TESLA-500 machine designs; the only solution may be a reduction in the bunch charge, and thus the luminosity, to reduce both the kink instability and the energy-z correlation for these measurements.

\begin{table} [tbp]
\caption{Summary of $\ecmbias$, due to energy spread and kink instability, at NLC-500 and TESLA-500 for both \epem and \emem collider modes.}
\vspace{2mm}

{\begin{tabular}{@{}|c|c|c|c|c|@{}} 

\hline
{\bf LC Machine Design}	& {\bf Collider Mode} & $\ecmbias$ & $\sigma(\ecmbias)$ & Max($\ecmbias$)	\\
			&		& $(\Delta y = 0)$ & $(\Delta y = 0)$ &   \\
\hline
NLC-500 	& \epem	& +520 ppm	& 170 ppm	& +1000 ppm			\\
NLC-500	& \emem	& +710 ppm	& 400 ppm	& +2000 ppm 		\\
TESLA-500	& \epem 	&  +50 ppm	& 30 ppm	& +250 ppm			\\
TESLA-500	& \emem	&  +230 ppm	& 120 ppm	& +500 ppm			\\
\hline
\end{tabular}}
\label{tab:ecmbias}
\end{table}

\newpage
\section{Polarization Studies}

The NLC extraction line design\cite{nosochkov1,nosochkov2} takes advantage of the large 20-mrad crossing angle and allows capability for beam diagnostics, including a Compton polarimeter.  Extraction line beam diagnostics are highly desirable at the LC.   There is much more flexibility in the beam optics design downstream of the IP compared to upstream to accommodate beam diagnostics.  One need not worry about emittance dilution or creating backgrounds that are problematic for the machine protection system or for the LC Detector at the IP.  The energy spectrometer and polarimeter can be closer to the LC IP with less extrapolation error from their measurements to the relevant beam quantities at the IP.  Given the high precision desired for both energy and polarization measurements, it is also very desirable to have redundant measurements of these quantities by independent techniques.  Extraction line diagnostics are needed to provide this.  For beam energy measurements, we plan to implement both an upstream BPM energy spectrometer (as was done for LEP-II) and a downstream extraction line synchrotron stripe energy spectrometer (as was done at SLC).  An extraction line polarimeter measurement can be compared to an upstream polarimeter measurement.  In addition, the extraction line more easily accommodates a back-scattered Compton gamma measurement to complement the back-scattered Compton electron measurement.  Beam-beam collision effects can be directly measured with extraction line diagnostics by comparing measurements with and without collisions.  The extraction line environment is difficult, though, due to the disrupted primary beams and secondary particles from collisions, especially the intense beamsstrahlung photons.  It is therefore necessary to demonstrate that meaningful extraction line diagnostics are feasible.  Here we present a feasible extraction line polarimeter design which has much more capability than a polarimeter upstream of the LC IP.

\begin{figure}
\begin{center}
\epsfig{file=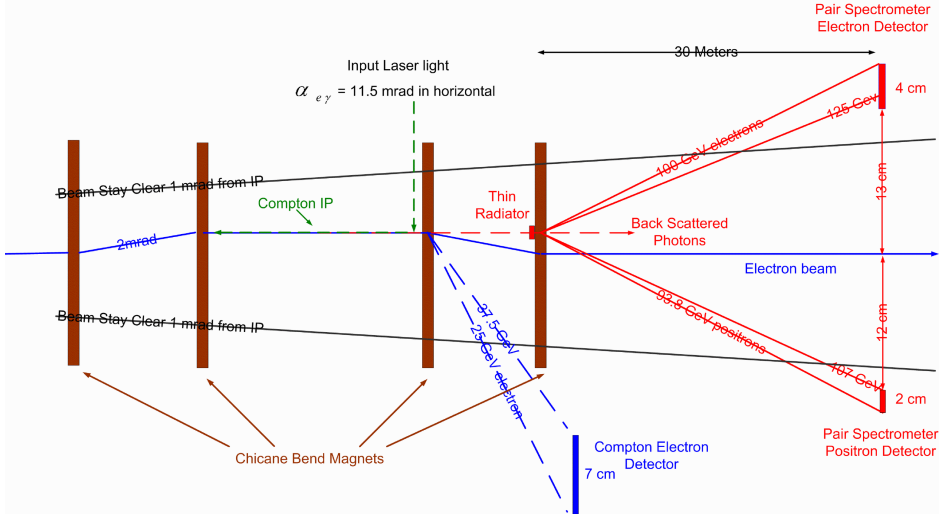,width=12.4cm}
\end{center}
\caption{One design option being studied for the extraction line chicane, with a Compton IP at mid-chicane for polarimetry.}
\label{fig:Compton_chicane}
\end{figure}

The NLC extraction line design achieves an 80$\%$ energy bandpass for transporting the disrupted electron beam to the beam dump.\cite{nosochkov2}  In this study we simulate transport of (realistic) disrupted NLC-500 beams, using the input TRC files described in the previous section, from the IP to the Compton IP.  Less than 0.3$\%$ of the disrupted beam particles are lost.  These losses should be tolerable for extraction line beam diagnostics, if suitable collimators and detector shielding are implemented.

The NLC group is working on detailed designs for polarization and energy measurements in the extraction line.\cite{ipbiwhite,ipbi}  Extraction line polarimetry is feasible for both \epem and \emem collider modes and we present some studies for this below.  The energy spectrometer design study is ongoing, including its impact on polarimetry.  One concept for the extraction line chicane that could be used for both a polarimeter and an energy spectrometer is shown in Figure~\ref{fig:Compton_chicane}.  

For polarimetery, a 532-nm circularly polarized laser beam collides with the electron beam in the middle of a vertical chicane with a horizontal crossing angle of 11.5 mrad.\cite{moffeit}  The laser pulse energy at the Compton IP is 100 mJ in a 2-ns FWHM pulse.  Compton-scattered electrons near the kinematic edge at 25.1 GeV are detected in a segmented detector, and the scattering rate asymmetry for electron and photon spins aligned versus anti-aligned can be used to determine the beam polarization.\cite{woods1}  We are also investigating the possibility of a pair spectrometer to measure the beam polarization from a (counting mode) measurement of the back-scattered Compton gamma asymmetry.  The detectors for back-scattered Compton electrons and gammas are outside of a 1-mrad stayclear, which is needed to accommodate the intense beamsstrahlung photon flux.  The converter for the Compton gammas is inside the 1-mrad stayclear, though, and this is likely only possible during dedicated polarimetry studies with no collisions.

\begin{figure}
\begin{center}
\epsfig{file=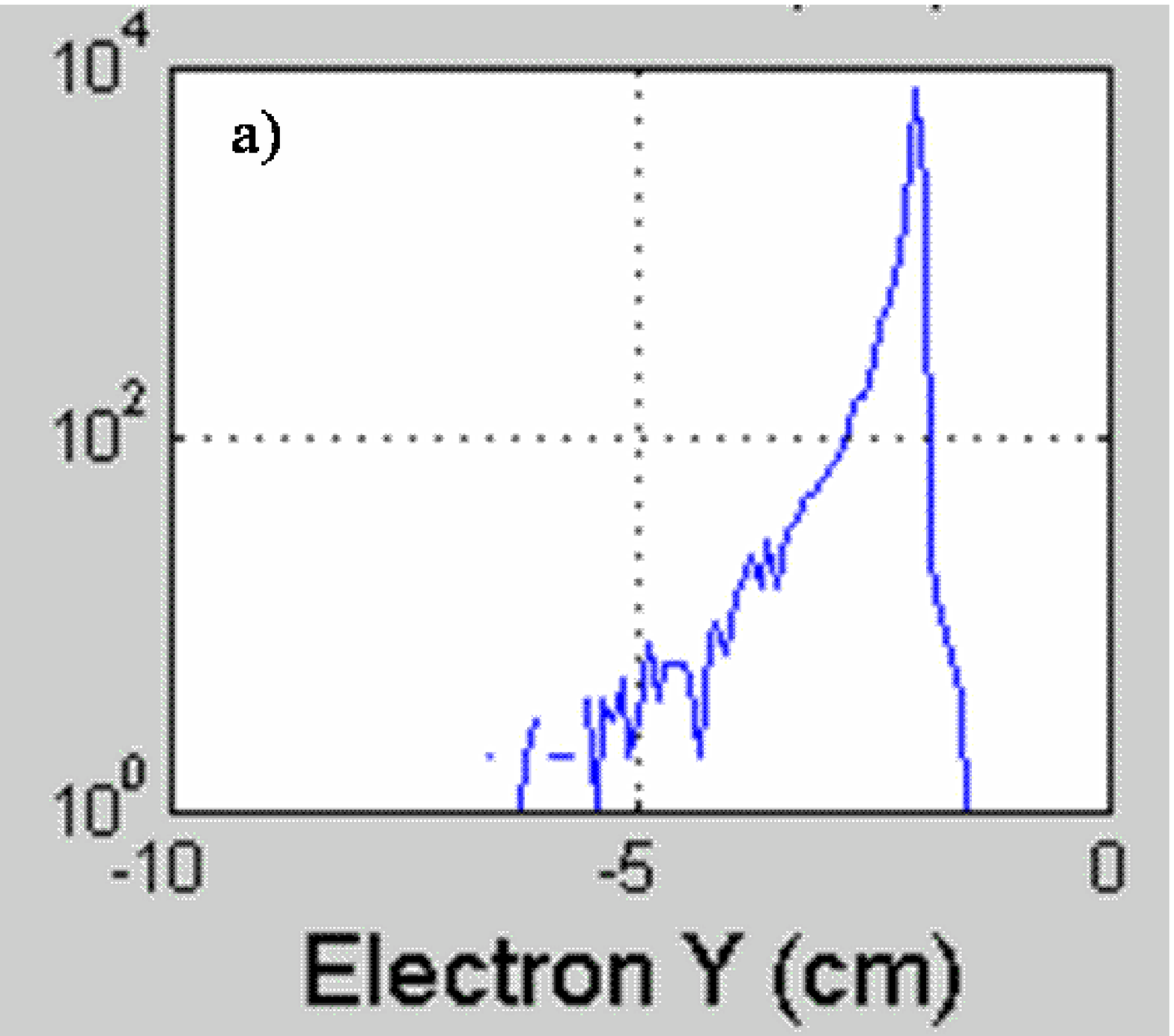,width=6.2cm}
\epsfig{file=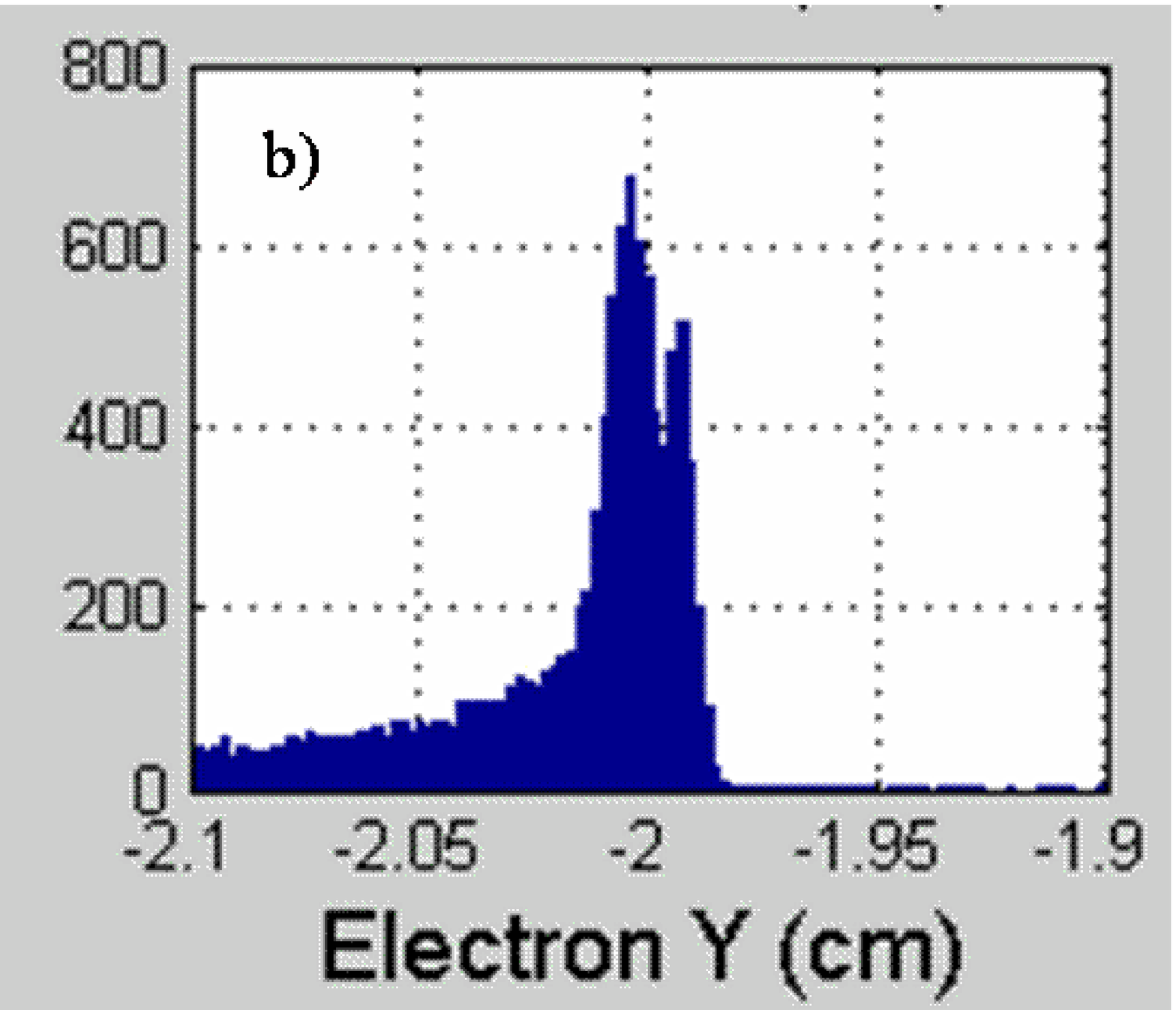,width=6.2cm}
\end{center}
\caption{The disrupted vertical beam distribution at the Compton IP for NLC-500 \epem collisions:  a) full distribution on a log scale and b) zooming in on the core of the beam} 
\label{fig:ypos_cip}
\end{figure}

Using the existing extraction line design,\cite{nosochkov2} whose chicane differs slightly from that shown in Figure~\ref{fig:Compton_chicane}, we use a GEANT-3 simulation to transport beams from the LC IP to the Compton IP and to determine the disrupted beam profile there.  The vertical profile of the disrupted electron beam (for NLC-500 \epem collisions) at the Compton IP is shown in Figure~\ref{fig:ypos_cip}.  The vertical dispersion of 20 mm and the beam energy spread are responsible for the double-peaked structure visible at $y=-2$cm in Figure~\ref{fig:ypos_cip}b).  The laser beam at the Compton IP is expected to have an rms width of $\approx 100 \mu$m, which is roughly matched to the size of the core of the disrupted electron beam.

\begin{figure}
\begin{center}
\epsfig{file=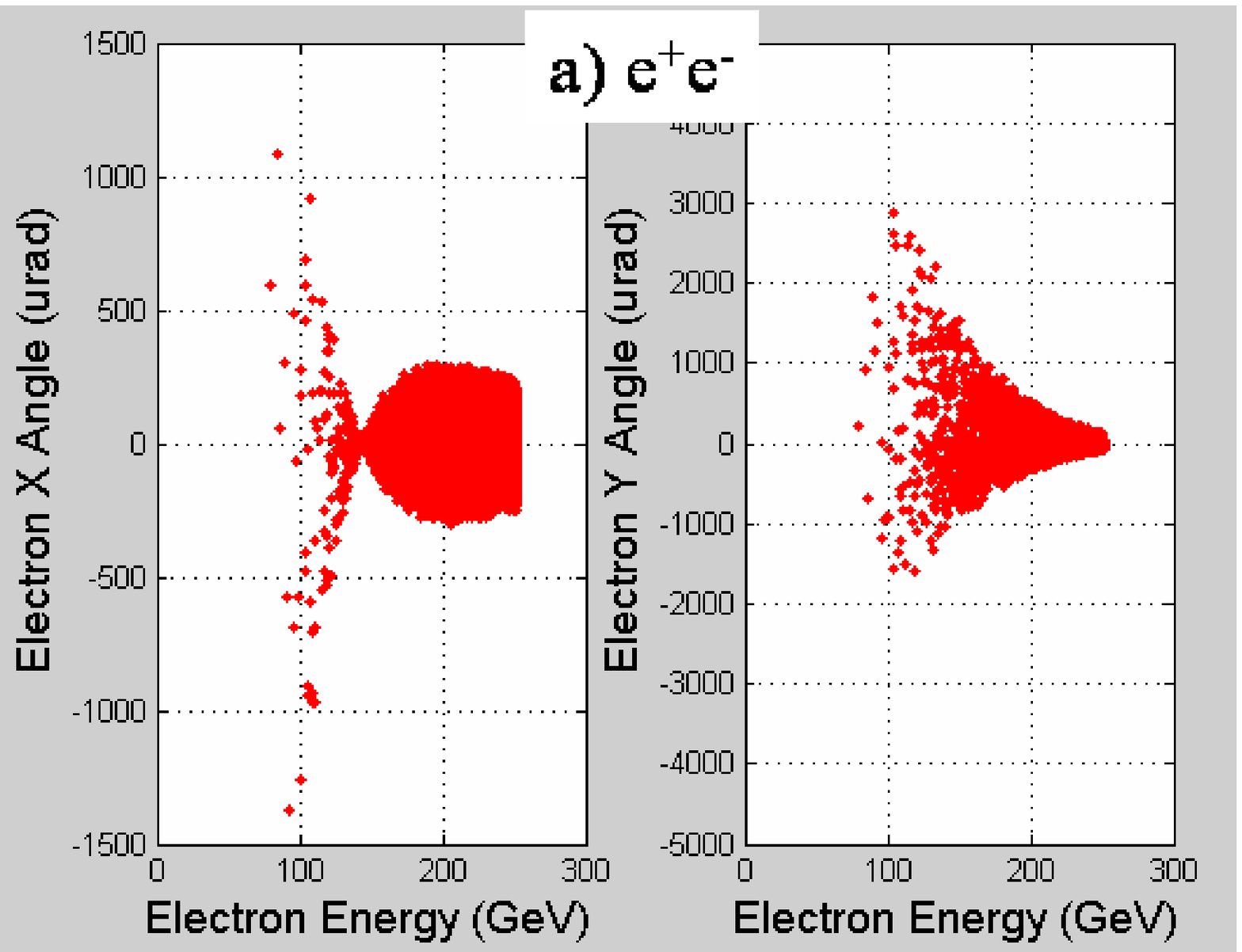,width=6.2cm}
\epsfig{file=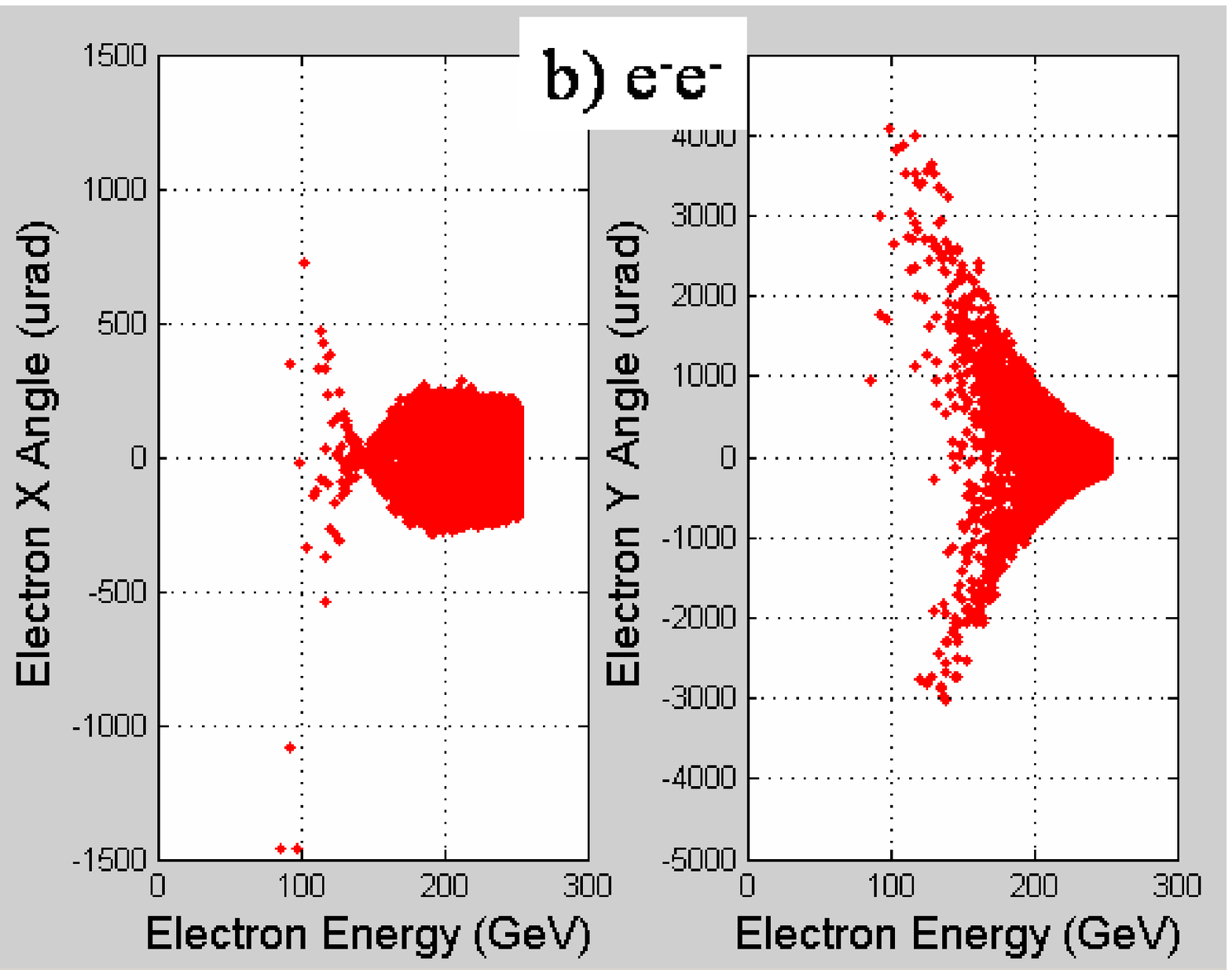,width=6.2cm}
\end{center}
\caption{The disrupted beam angles at the Compton IP for a) \epem and b) \emem collisions, for one set of NLC-500 TRC files.}
\label{fig:CIPangle-enlc}
\end{figure}

\begin{figure}
\begin{center}
\epsfig{file=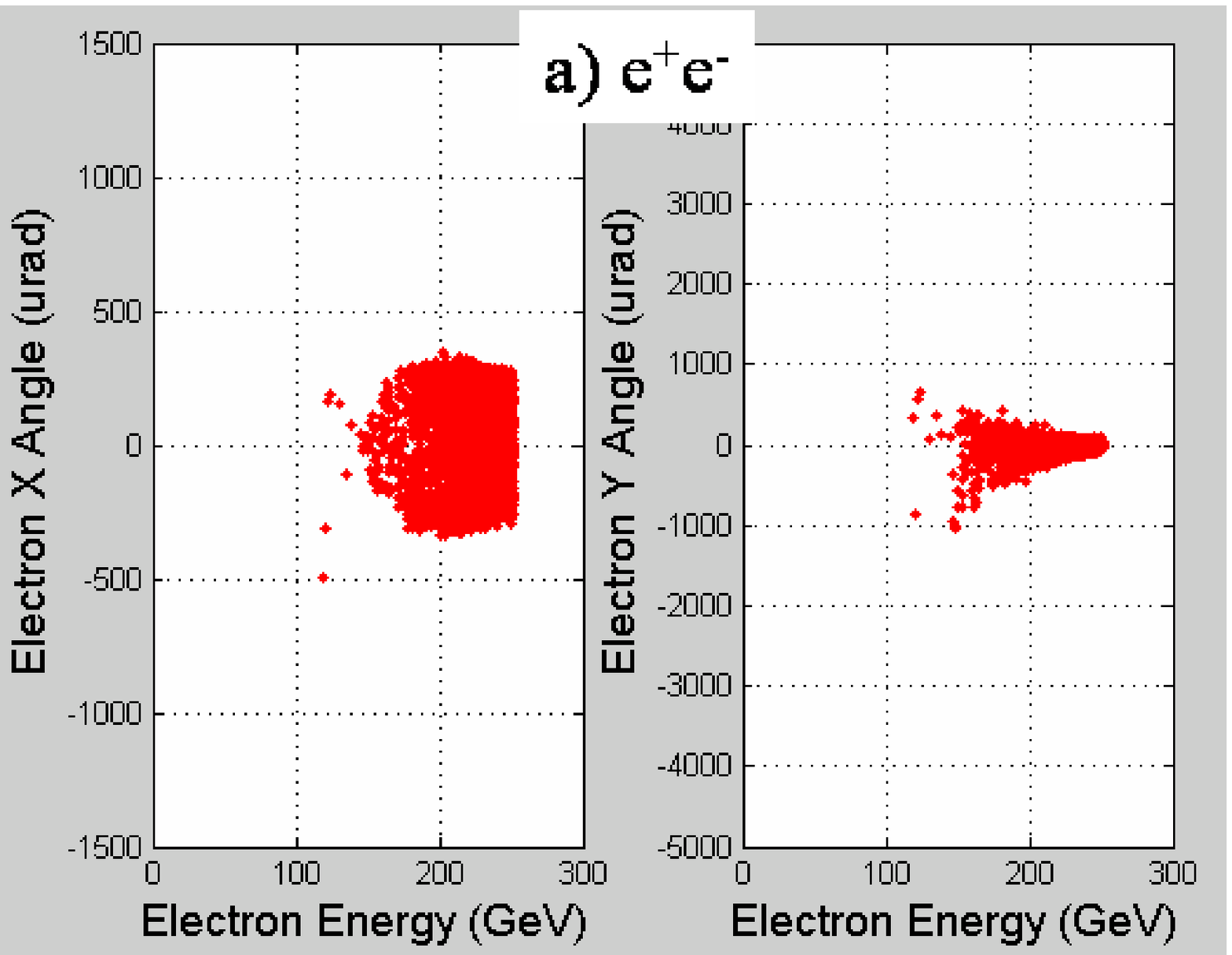,width=6.2cm}
\epsfig{file=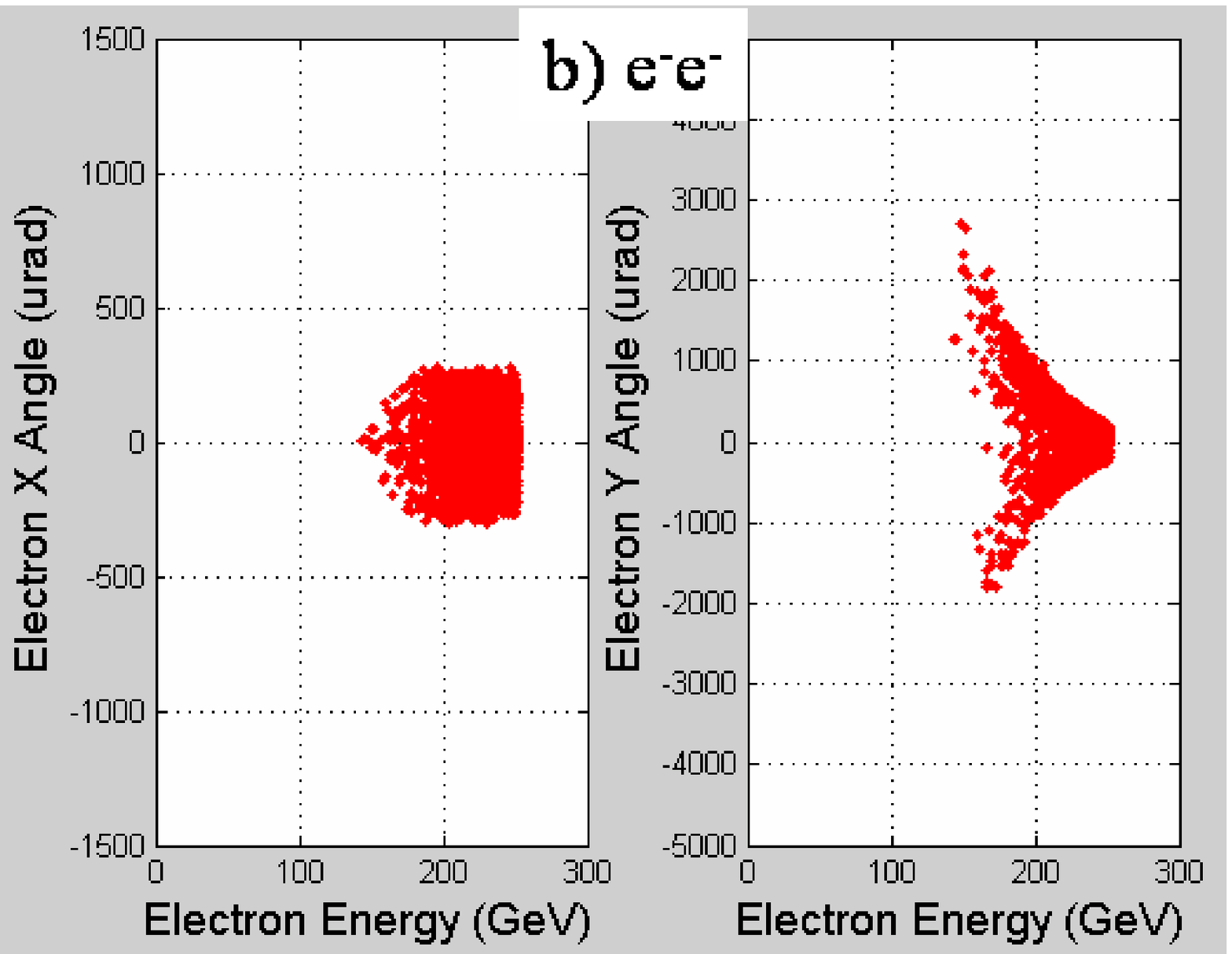,width=6.2cm}
\end{center}
\caption{The disrupted beam angles at the Compton IP for a) \epem and b) \emem collisions, for one set of TESLA-500 TRC files.}
\label{fig:CIPangle-etesla}
\end{figure}

The angular distributions of the disrupted electron beam at the Compton IP are shown in Figures~\ref{fig:CIPangle-enlc} and~\ref{fig:CIPangle-etesla} for both \epem and \emem collider modes.  Above 225 GeV ($90\%$ of the beam energy) the angular distributions are well behaved for both \epem and \emem collider modes at NLC and TESLA.  In addition to providing polarization measurements for disrupted electrons near the endpoint beam energy, meaningful polarization measurements may be achievable for disrupted electrons in the region from 225-250 GeV.  Angular spread in the electron beam leads to a spread in the spin precession (spin diffusion) and an effective depolarization.  Spin precession and depolarization and their impact on polarimetry is discussed further in Appendices A and B.
Table~\ref{tab:deltap} summarizes parameters for the disrupted beams in the extraction line, including beam losses in transport to the mid-chicane point for the NLC extraction line.  Beam losses are well below $0.1\%$ except for \emem collisions at NLC-500, where they are $\approx 0.3\%$.  

\begin{table} [tbp]
\caption{Extraction Line Beam Properities:  angular divergences of the disrupted beams at the Linear Collider IP; luminosity-weighted depolarization; average beamstrahlung energy loss; and $\%$ beam loss in transport from the IP to mid-chicane of the NLC extraction line.  Results are averaged over 6 TRC files.}
\vspace{2mm}

{\begin{tabular}{@{}|c|c|c|c|c|@{}} 
\hline
{\bf Parameter}	& {\bf NLC-500} 	& {\bf TESLA-500}	& {\bf NLC-500} 	& {\bf TESLA-500}	\\
	& \epem	& \epem	& \emem	& \emem	\\
\hline
$\sigma (\theta_x)$ & 228 $\mu$rad	& 275 $\mu$rad	& 182 $\mu$rad			& 198 $\mu$rad	\\
$\sigma (\theta_y)$ & 85 $\mu$rad	& 56 $\mu$rad	& 185 $\mu$rad			& 236 $\mu$rad	\\
$\dplum$		& $0.24\%$		& $0.32\%$		& $0.36\%$				& $0.45\%$ \\
$\Delta E$		& $6.2\%$		& $4.2\%$		& $5.2\%$			
	& $3.5\%$	\\
Chicane losses	& $<0.1\%$		& $<0.1\%$		& $0.3\%$
	& $<0.1\%$	\\
\hline
\end{tabular}}
\label{tab:deltap}
\end{table}

\begin{figure}
\begin{center}
\epsfig{file=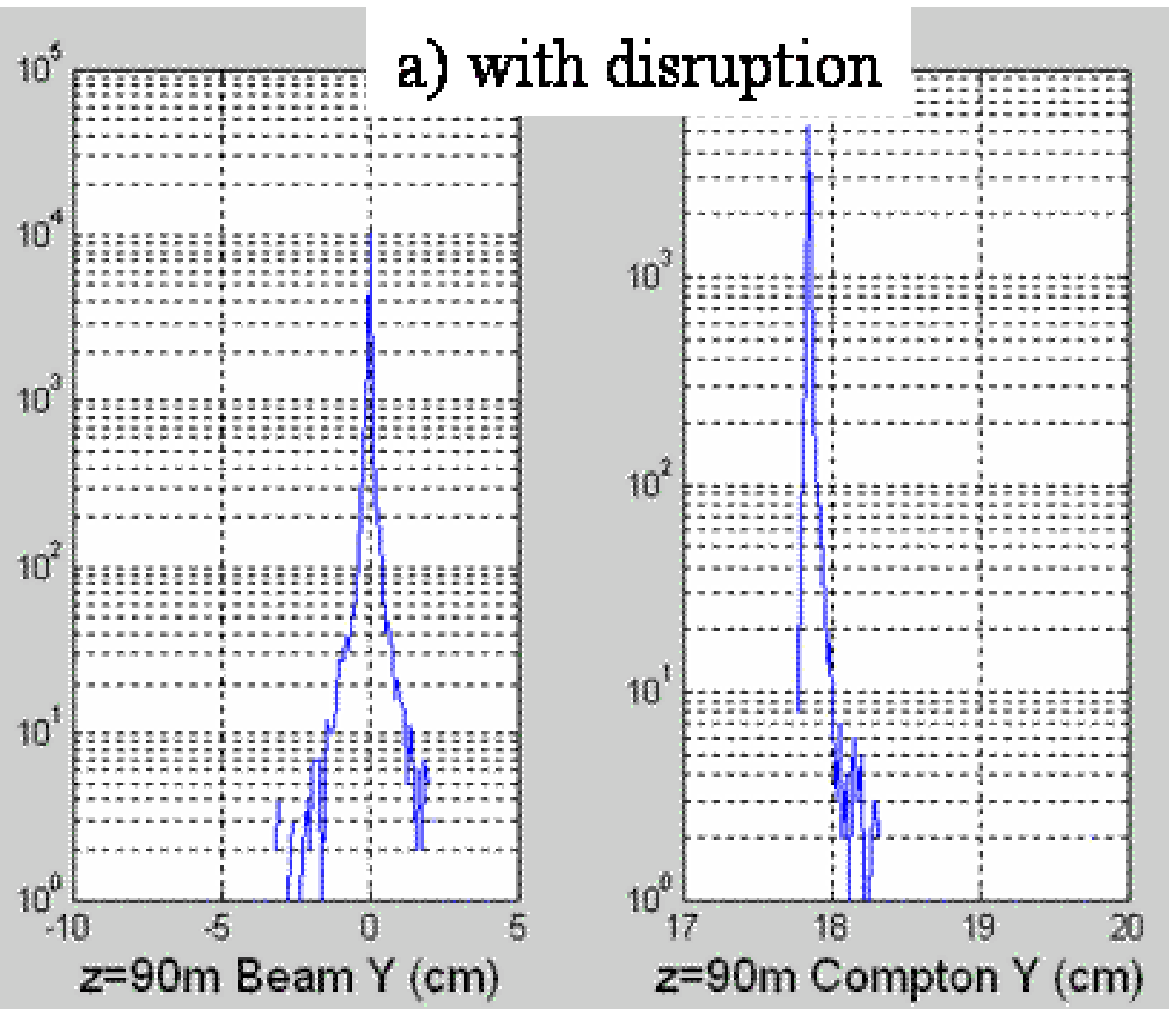,width=6.2cm}
\epsfig{file=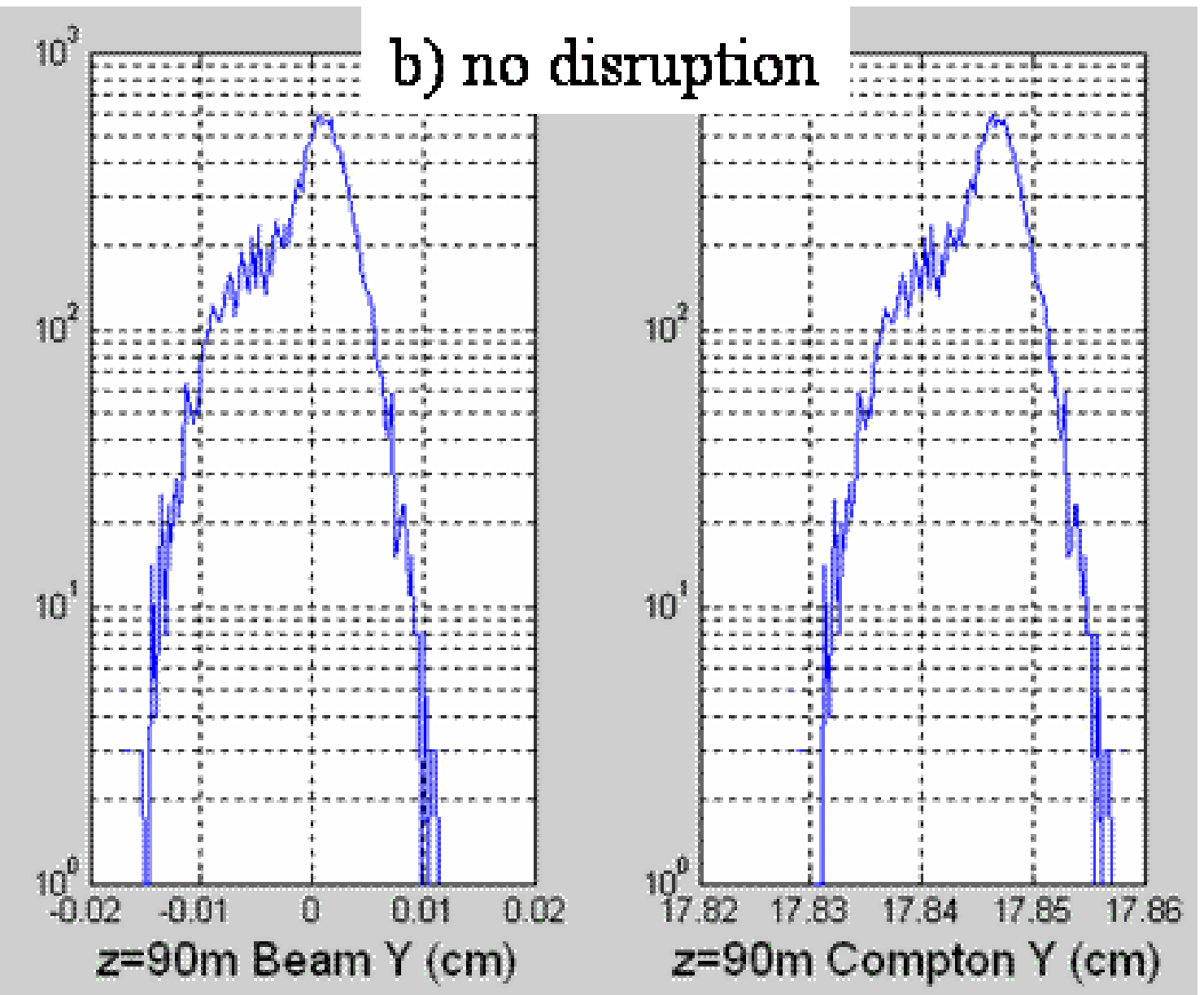,width=6.2cm}
\end{center}
\caption{The vertical beam distribution at the Compton detector plane for the NLC-500 \epem collider mode, a) with collisions (ie. disrupted outgoing beam) and b) without collisions (no disruption).} 
\label{fig:z90-ypos}
\end{figure}

We have simulated the transport of Compton-scattered electrons at the endpoint energy of 25.1 GeV to a detector plane 90 meters downstream of the LC IP, to investigate the separation of the Compton signal from the disrupted electron beam distribution and from the beamsstrahlung photons.  Distributions for the Compton endpoint electrons and the beam electrons at this location are shown in Figure~\ref{fig:z90-ypos} for one set of NLC-500 files with \epem collisions, both with and without disruption.  Even with collisions, there is very good separation between the Compton signal and the disrupted electron beam.  The Compton endpoint is 18 cm from the beam axis and is well outside the 1-mrad stayclear of 9 cm.  The expected Compton scattering rate at the 25.1 GeV endpoint is 500 Compton electrons per GeV (or 600 Compton electrons per cm at the detector plane) per pulse for the NLC-500 design.\cite{moffeit}

\newpage
\section{Summary}

We have presented results from luminosity, energy and polarization studies for a 500 GeV Linear Collider.  We compared NLC and TESLA beam parameters for both \epem and \emem modes of operation, using realistic colliding beam distributions.  We find that the narrow deflection scan for \emem collisions creates significant difficulties for a beam-based feedback to stabilize collisions, and that a realistic \emem luminosity may only be $10-20\%$ of the \epem luminosity.  We find that the combined effect of beam energy spread, energy-z correlations within a bunch, and a kink instability result in a significant difference between the average $\ecm$ measured by energy spectrometers and the luminosity-weighted $\ecm$.  This difference results in a bias, $\ecmbias$, if the luminosity spectrum is being determined from energy spectrometer and Bhabha acollinearity measurements.  $\ecmbias$ is a factor two larger for \emem collisions than for \epem collisions, and it is a factor 3-10 larger for the NLC beam parameters than for the TESLA beam parameters.  New analyses, utilizing $\gamma  Z, ZZ$ and $WW$ events and making use of existing $Z$ and $W$ mass measurements, are needed to provide more robust determinations of $\ecmbias$; they should be able to achieve less than 200 ppm uncertainty on $\ecmlum$ for both NLC and TESLA designs at nominal luminosity.   But this {\it kink instability} effect may prevent achieving 50 ppm precision or better on $\ecmlum$ (for both NLC and TESLA designs), which is desired for improved W mass or (ultimate) Giga-Z \alr measurements.  For polarimetry, we have shown that an extraction line Compton polarimeter is feasible, if the collider IP has a 20-mrad crossing angle.  It has much more capability than an upstream polarimeter to achieve the best precision for determining the luminosity-weighted beam polarization.

\newpage
\appendix{.  Depolarization effects and Polarimetry}

(BMT) Spin precession with respect to the electron momentum vector is given by
\begin{equation}
\theta_{spin} = \frac{E(GeV)}{0.44065} \cdot \theta_{bend}.
\end{equation}
Depolarization resulting from spin diffusion is given by
\begin{equation}
\dpBMT = 1-\cos \left[ \sigma \left( \theta_{spin} \right) \right],
\end{equation}
where $\sigma (\theta_{spin} )$ is the rms of the spin precession distribution.
At the IP, the incoming beams have small enough angular divergence $(<50 \mu$rad rms) that the resulting $\dpIPBMT$ is negligible.  However, the outgoing angular divergences are significantly larger (see Table~\ref{tab:deltap}), which results in $\dpIPBMT \approx (1.0-1.8) \%$ for the outgoing beams to the extraction line.  
Depolarization of the outgoing beams at the IP can result both from BMT spin precession and from Sokolov-Ternov spin flips from beam-beam effects,\cite{yokoya}
\begin{equation}
\dpIP=\dpIPBMT +\dpIPST
\end{equation}
The luminosity-weighted depolarization is smaller, however, and is typically about $1/4$ of the outgoing beam depolarization\cite{thompson} (whether due to BMT spin precession or due to ST spin flips),
\begin{eqnarray}
\dplum & = & \dplumBMT + \dplumST \\
\dplum & \approx & \frac{1}{4} (\dpIPBMT + \dpIPST)
\end{eqnarray}

The Compton polarimeter in the extraction line can measure the difference in polarization between collisions and no collisions, which has contributions from both BMT spin precession and Sokolov-Ternov spin flips:
\begin{equation}
\dpCIPmeas = \dpCIPBMT + \dpIPST.
\end{equation}
$\dpCIPBMT$ can differ from $\dpIPBMT$ due to the R-Transport matrix from the IP to the Compton IP.  This matrix should be well known from the beam optics.  (The current NLC extraction line has an angular magnification from the IP to the CIP of $\approx 0.5$, so that $\dpCIPBMT \approx \dplumBMT$.)  If the disrupted beam angles can be determined (from simulations or from extraction line measurements) then one can infer both $\dpIPBMT$ and $\dpCIPBMT$.  The extraction line measurement of $\dpCIPmeas$ can then be used to determine $\dpIPST$.  

As an example, we find for one of the (TRC) NLC-500 files that \epem collisions give $\dplumBMT = 0.22\%$.  The R-Transport matrix to the Compton IP then predicts that $\dpCIPBMT = 0.31\%$.  Using the GEANT-3 simulation to transport the disrupted electron beam to the CIP and applying a weighting for the collision with the Compton laser beam we would expect to measure $\dpCIPBMT = 0.28\%$, in very good agreement with the predicted value.  For these NLC-500 beam parameters we expect $\dpIPST \approx 0.4\%$ and $\dplumST \approx 0.1\%$.\cite{thompson}

\newpage
\appendix{.  Spin Precession from an IP Crossing Angle}
In the NLC-500 design, the beams collide at the IP with a 20-mrad crossing angle.  The beam trajectories are therefore mis-aligned by 10 mrad with the detector solenoid field.  This results in a vertical kick to the beams.  The vertical kick from the solenoid in the barrel region of the LC detector is partially cancelled by a compensating kick in the endcap fringe field.  In one study of this effect for NLC-500,\cite{tenenbaum} the deflection angle with respect to the incoming beam trajectory would be 68 $\mu$rad at the LC IP and 136 $\mu$rad for the outgoing beam to the extraction line.  These deflection angles are energy dependent,
\begin{eqnarray}
\theta_y^{IP} & = & 68 \mu rad \cdot \left( \frac{250 GeV}{E} \right) \\
\theta_y^{extraction} & = & 136 \mu rad \cdot \left( \frac{250 GeV}{E} \right). 
\end{eqnarray}

Three problems result if these vertical kicks are not compensated:
\begin{itemize}
\item The extraction line must be realigned when the beam energy is changed.
\item There will be a 136 $\mu$rad vertical crossing angle for \emem collisions.  (\epem still collide head-on, but at an angle of 68$\mu$rad with respect to the incoming beam trajectories or the solenoid field.)  This reduces the luminosity to nearly zero.
\item There is a net band angle of 68 $\mu$rad between the beam trajectory at the (upstream or downstream) polarimeters and the IP.  This angle is small compared to the rms divergence of the disrupted beams, but nonetheless is undesirable.
\end{itemize}
Given the above problems, it is clear that the vertical kicks from the solenoid in the crossing angle geometry must be compensated.  It should be straight forward to compensate for the vertical kick downstream of the IP.  More care is needed for compensating the upstream kick because of the stringent requirements for emittance preservation and beam alignment.  Design studies are in progress and should include a solution with no net bend angles between the Linear Collider IP and the polarimeter Compton IP. 

\nonumsection{References}

\end{document}